\documentclass[aps,prb,superscriptaddress,twocolumn,a4paper]{revtex4-1}
\usepackage{amsmath,empheq}
\usepackage{amssymb}
\usepackage{mathrsfs}  
\usepackage{amsfonts}
\usepackage{booktabs}
\usepackage{physics}
\usepackage{graphicx} 
\usepackage{subfigure} 
\usepackage{epsfig}
\usepackage{color}
\usepackage{epstopdf}
\usepackage{xspace}
 \usepackage{rotating}
 \usepackage{appendix}
 \usepackage{chemformula}
 \usepackage{bm}
 \usepackage[colorlinks=true,linkcolor=blue,citecolor=blue,filecolor=blue,urlcolor=blue]{hyperref}
\definecolor{bondiblue}{rgb}{0.0, 0.58, 0.71}

\newcommand{\ua}{\uparrow}
\newcommand{\da}{\downarrow}

\begin{document}

\title{Even-odd effect in multilayer Kitaev honeycomb magnets}
\author{Jaime Merino}
\affiliation{Departamento de F\'isica Te\'orica de la Materia Condensada, Condensed Matter Physics Center (IFIMAC) and
Instituto Nicol\'as Cabrera, Universidad Aut\'onoma de Madrid, Madrid 28049, Spain}
\author{Arnaud Ralko}
\affiliation{Institut N\'eel, UPR2940, Universit\'e Grenoble Alpes et CNRS, Grenoble 38042, France}

\begin{abstract}
Motivated by the three-dimensional structure of Kitaev materials we explore multilayer Kitaev models. The magnetic properties of a multilayer of an arbitrary number of Kitaev honeycomb layers stacked on top of each other coupled through a Heisenberg interaction, $J$, is analyzed through Abrikosov fermion mean-field theory. The system sustains quantum spin liquid (QSL) solutions which have different character depending on parity of the number of layers. While in even layered Kitaev models a gapped QSL emerges, odd-layered models host gapless QSLs. The projective symmetry group analysis of these solutions unravel a layer-to-layer inversion symmetry rather than an expected reflection. 
Although these QSLs retain features of the single layer Kitaev spin liquid (KSL), they should be regarded hybrid QSLs consisting on several KSLs. The good agreement at large-$J$ between the energy of the Gutzwiller projected mean-field QSL and the exact energy indicates that such QSL {\it Ansatz} is adiabatically connected to the exact ground state. We also find that the Kitaev gapped chiral quantum spin liquid induced by external magnetic fields is stabilized by an antiferromagnetic interlayer coupling. Our results are relevant to the physics of $\alpha$-RuCl$_3$ and H$_3$LiIr$_2$O$_6$ which are examples of magnetically 
coupled multilayer Kitaev models. 
\end{abstract}
 \date{\today}
 \maketitle
\section{Introduction} 

The Kitaev spin liquid (KSL), a quantum spin liquid on a honeycomb quantum compass model, has attracted a lot of attention since its possible realization in candidate materials such as  $\alpha$-RuCl$_3$. In spite of its quasi-2D structure   there is significant interlayer magnetic exchange couplings between Kitaev honeycomb layers \cite{slagle_theory_2018} whose role is yet poorly understood. Majorana mean-field theory on the antiferromagnetic (AFM) coupled bilayer Kitaev model \cite{seifert_bilayer_2018} predicts an intriguing QSL with non-zero interlayer flux which is not present when the simpler Ising interlayer coupling exists\cite{vijayvargia_topological_2024}. Exact diagonalization (ED) on the bilayer Kitaev 
model finds a quantum phase transition between the KSL and interlayer dimer singlet state for AF interlayer exchange coupling, and a smooth crossover to another QSL
emerges in the $S=1$ Kitaev model with increasing ferromagnetic (FM) interlayer
coupling\cite{tomishige_low-temperature_2019,tomishige_interlayer_2018}. Most of these studies focus on models consisting of only two Kitaev layers. In weakly coupled Kitaev multilayers, the chiral KSL \cite{joy_gauge_2024} is found to be 
stable in a finite range of FM/AFM interlayer couplings. These preliminary works raise important questions: do Kitaev models consisting on two or more layers host QSLs?, if these exist, what is their nature?, are they gapped or ungapped?, how do their properties depend on the number of layers?.

\begin{figure}[h!]
    \centering
 \includegraphics[width=8cm,clip]{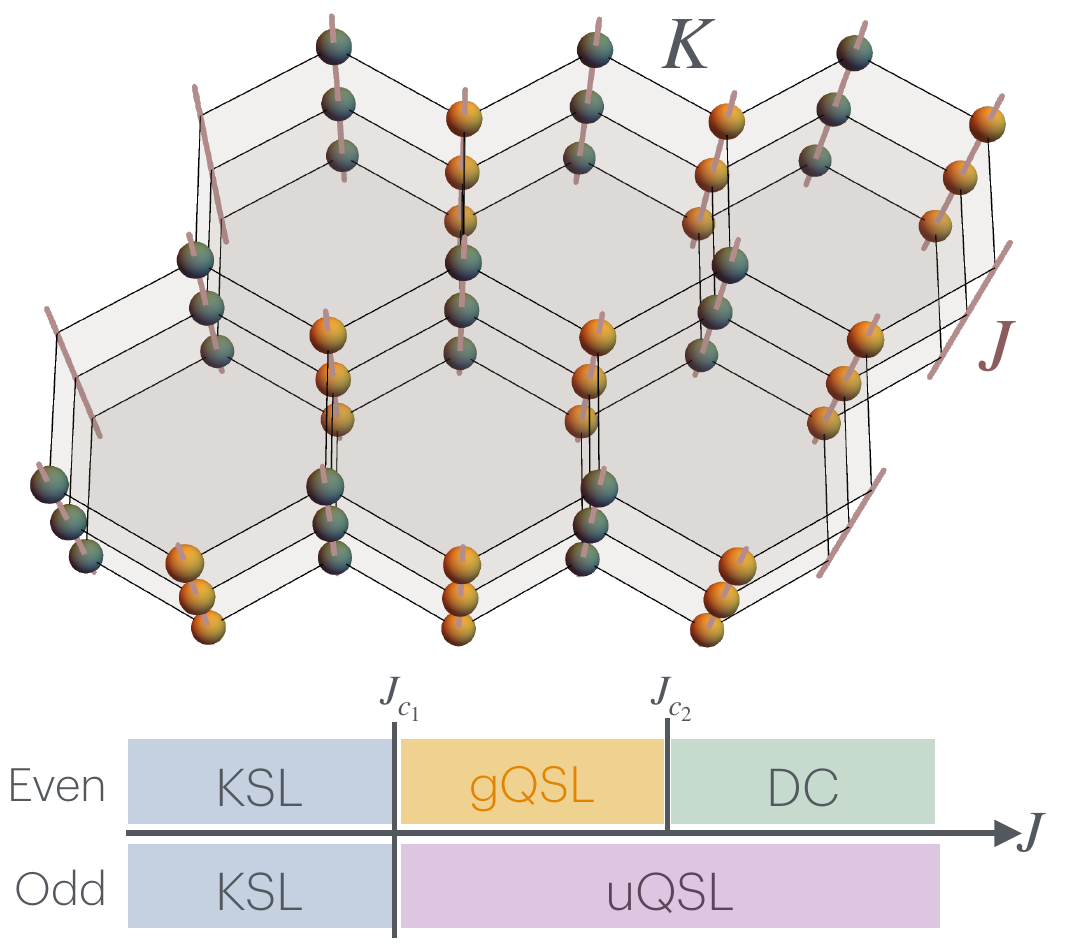} 
    \caption{Generic phase diagrams of multilayered Kitaev models, illustrated on the top panel with the trilayer case. While ungapped QSLs (uQSL) arise in odd-layered Kitaev models, even-layered Kitaev models host gapped QSLs (gQSL) at intermediate $J$. In the even-layered model a decoupled crystal (DC) phase is stabilized at large interlayer couplings $J>J_{c2}$. The KSL denotes the conventional Kitaev QSL on decoupled layers. The phase diagram is qualitatively similar for FM, $J>0$, 
    and AFM, $J<0$, systems with only $J_{c1}, J_{c2}$ values changing. In the odd case, the DC would also occur but only at $J=\infty$  . For FM 
    $J \rightarrow \infty $, the model maps onto a honeycomb lattice with $S_{tot} = n/2$ where $n$ is the total number of layers. The Kitaev coupling is fixed to $K=1$ in the lower plot.
     }
\label{fig:structure}
\end{figure}

Multilayer Kitaev models with strong FM coupling map onto $S>{1 \over 2}$ 
2D Kitaev model on a honeycomb lattice. 
A conjecture similar to Haldane's conjecture on AF Heisenberg spin chains\cite{haldane_continuum_1983} states that the integer(half-integer) Kitaev model on the honeycomb lattice hosts a gapped (gapless) quantum spin liquid \cite{ma_mathbbz_2_2023}. 
However, since there is no exact solution to such model, as for the $S=1/2$ Kitaev model, numerical methods are needed to verify this conjecture.
It is consistent with tensor network calculations which suggest the existence of a gapped \cite{dong_spin-1_2020,khait_characterizing_2021,chen_excitation_2022}, $\mathbb{Z}_2$ QSL with bosonic excitations 
in the $S=1$ Kitaev model on the honeycomb lattice.  
Recent numerical work \cite{pohle_eight-color_2024,georgiou_spin-s_2024} finds 
finite regions of QSL behavior in the phase diagram of the $S=1$ 
Heisenberg-Kitaev model. 
Apart from its theoretical interest the $S>1/2$, Kitaev model may be 
realized in several materials. While A$_3$Ni$_2$XO$_6$ (A=Li, Na and X=Bi, Sb), 
Na$_2$Ni$_2$TeO$_6$ \cite{samarakoon_comprehensive_2017}, and KNiAsO$_4$ \cite{taddei_zigzag_2023} are considered candidates for the $S=1$ Heisenberg-Kitaev model, the $S=3/2$  Kitaev model is relevant to CrI$_3$,CrGeTe$_3$\cite{xu_interplay_2018}  
monolayers and  CrSiTe$_3$.\cite{xu_possible_2020}

In the present work, we analyze possible QSLs emerging in the multilayered Kitaev systems shown in Fig.~\ref{fig:structure} and described by model (\ref{eq:model}). A key result of our work is also summarized in Fig.~\ref{fig:structure} which displays the phase diagram of multilayered Kitaev models using an Abrikosov parton fermion construction\cite{choi_topological_2018}. The figure shows how QSLs solutions in even and odd-layered Kitaev models are qualitatively different. While a gapped QSL  (gQSL) arises at intermediate interlayer coupling $J$ in even-layered systems, in odd-layered cases an ungapped QSL (uQSL) emerges.
In the even-layered case a decoupled crystal (DC) phase consisting on a set of disconnected interlayer dimers ($2N$ dimers corresponding to $N$ unit cells) between consecutive layers emerges. The dimers consists on spin singlets or triplets depending on whether $J$ is FM ($J>0$) or AFM ($J<0$), respectively. Our results support  Ma's \cite{ma_mathbbz_2_2023} conjecture by which while the integer $S$ Kitaev model can host a gQSL, the half-odd integer $S$ model hosts an uQSL. Despite its small size limitations, an ED analysis of the multilayer Kitaev model allows to validate our mean-field QSL solution which is found to be adiabatically connected to the exact ground state. We finally find that the critical magnetic 
field to the fully polarized (FP) phase is suppressed (enhanced) 
by a FM (AFM) interlayer coupling. Following previous projective symmetry group analysis on the single layer\cite{you_doping_2012} KSL, we have considered the case of the bilayer KSL and showed that it is only compatible with a specific interlayer inversion symmetry structure rather than an expected reflection, where two opposite copies of the single layer KSL can be coupled together. Our results are relevant to Kitaev layered materials such as $\alpha$-RuCl$_3$ and the iridate compound, H$_3$LiIr$_2$O$_6$. 

The rest of the paper is organized as follows. In Sec.~\ref{sec:model}
we introduce the multilayered Kitaev model and the Abrikosov parton fermion approach for analyzing it. In Sec.~\ref{sec:qsl} we discuss the quantum spin liquid solutions 
and the generic magnetic properties arising in the model.  In Sec.~\ref{sec:interlayer} we provide an analysis of interlayer magnetic correlations which goes beyond mean-field theory and compared to ED procedures. Sec.~\ref{sec:magnetic} is devoted to 
the effects of a magnetic field on FM and AFM coupled multilayers. Finally, in Sec.~\ref{sec:conclusions} we wrap up with some conclusions discussing the relevance of our work to actual materials.

\section{Model and methods} 
\label{sec:model}

We consider a set of an arbitrary number of honeycomb layers realizing a $S=1/2$ Kitaev model stacked on top of each other as depicted in Fig.~\ref{fig:structure}. The layers are coupled through a Heisenberg interaction such as:
\begin{equation}
H=H_K+H_J,
\label{eq:model}
\end{equation}
with respectively the Kitaev\cite{kitaev_anyons_2006} and Heisenberg parts 
\begin{eqnarray}
H_K&=& -K \sum_l  \sum_{\langle i ,j \rangle_a } \hat{S}^a_{l,i} \hat{S}^a_{l,j},
\nonumber \\
H_J &=&-J \sum_{i,l} \hat{\bf S}_{l,i}\cdot \hat{\bf S}_{l+1,j}.
\end{eqnarray}

In these expressions, $\langle i ,j \rangle_a$ denotes all intralayer nearest neighbor bonds having one of the three natural direction $a =x,y,z$, and   $l=1,2,\cdots,n$  is the layer index. We will denote by $N$ the number of unit cells of the system. In our definition, a $J>0$ ($J<0$) corresponds to the FM (AFM) case. 

\subsection{Abrikosov fermions}

In order to analyze the various ground states of the model we use the Abrikosov representation 
of a spin\cite{lee_doping_2006,lee_high_2007} at site $i$ and layer $l$:
\begin{equation}
\hat{\bf S}_{l,i}= {1 \over 2} f^\dagger_{l,i} \boldsymbol{\sigma} f_{l,i},
\end{equation}
with $f_{l,i}^\dagger = ( f_{i,\uparrow}^\dagger,f_{i,\downarrow}^\dagger)$ and $\boldsymbol{\sigma} = \sigma^x {\bf u}_x+\sigma^y {\bf u}_y+\sigma^z {\bf u}_z$ are the Pauli matrices. 
It is possible to introduce operators in the singlet and triplet particle-particle and particle-hole channels in order to rewrite the Hamiltonian.
The singlet channels expressed in terms of the 
Abrikosov fermions read:
\begin{eqnarray}
\hat{h}^0_{li,l'j} &=& f^\dagger_{l,i} f_{l',j}, 
\nonumber \\
\hat{p}^0_{li,l'j} &=& -i f_{l,i} \sigma^y f_{l',j},
\end{eqnarray}
while the triplet channels read: 
\begin{eqnarray}
\hat{h}^a_{li,l'j} &=& f^\dagger_{l,i} \sigma^a f_{l',j},  
\nonumber \\
\hat{p}^a_{li,l'j} &=& i f_{l,i} \sigma^y \sigma^a f_{l',j},
\end{eqnarray}
with $a=x,y,z$ the real-space components.

In terms of these operators, the interaction between spins can be expressed as\cite{choi_topological_2018}:
\begin{eqnarray}
\label{bil}
	\hat{S}_{l,i}^a \hat{S}_{l',j}^a = \hat{\bf h}^\dagger_{li,l'j} C^a \hat{\bf h}_{li,l'j} + \hat{\bf p}^\dagger_{li,l'j} C^a \hat{\bf p}_{li,l'j},
\end{eqnarray}
with $4\times4$ component diagonal matrices defined as $\text{diag}(C^a) = (-1,(-1)^{\delta_{a,1}},(-1)^{\delta_{a,2}},(-1)^{\delta_{a,3}})/2$ and $\delta$ the Kronecker symbol, and where we have introduced the hopping vector $\hat{\bf h} = ( \hat{h}^0,\hat{h}^x,\hat{h}^y,\hat{h}^z )$ and the pairing vector $\hat{\bf p} = ( \hat{p}^0,\hat{p}^x,\hat{p}^y,\hat{p}^z)$, omitting the subscripts.

Defined this way, the isotropic interaction is nothing else but the total of the three components of the spin bilinears with an extra factor of $1/4$ in order to recover the exact energy ground state \cite{choi_topological_2018} $T = \frac{1}{4} \sum_{a} C^a $ such that 
\begin{eqnarray}
\hat{\bf S}_{l,i} \cdot \hat{\bf S}_{l',j} &=& \hat{\bf h}^\dagger_{li,l'j} T \hat{\bf h}_{li,l'j} + \hat{\bf p}^\dagger_{li,l'j} T \hat{\bf p}_{li,l'j}.
\label{eq:inter}
\end{eqnarray}
with diag$(T)=(-3/8,1/8,1/8,1/8)$.
In the case of the interlayer Heisenberg interactions, we simply have $l \ne l'$ and $i=j$.

\subsection{Mean-field theory}

In order to study this Hamiltonian, and since the four 
 fermion interaction is very hard to treat exactly, we decoupled the 
singlet and triplet channels to extract a mean-field Hamiltonian.
Thanks to the expressions of the spin bilinears of Eq.~(\ref{bil}), this is readily achieved by proceeding to the mean-field decoupling
\begin{eqnarray*}
{\hat{\bf h}}_{li,l'j}^\dagger M {\hat{\bf h}}_{li,l'j} &\to&  {\hat{\bf h}}_{li,l'j}^\dagger M {{\bf h}}_{li,l'j} +  {{\bf h}}_{li,l'j}^* M {\hat{\bf h}}_{li,l'j} \nonumber \\ &-&  {{\bf h}}_{li,l'j}^*M {{\bf h}}_{li,l'j}, \nonumber \\ 
  {\hat{\bf p}}_{li,l'j}^\dagger M {\hat{\bf p}}_{li,l'j}  &\to& {\hat{\bf p}}_{li,l'j}^\dagger M {{\bf p}}_{li,l'j}  + {{\bf p}}_{li,l'j}^* M {\hat{\bf p}}_{li,l'j} \nonumber \\ &-& {{\bf p}}_{li,l'j}^* M {{\bf p}}_{li,l'j},
  \label{eq:hmf}
\end{eqnarray*}
where we have defined the expectation pairing and hopping mean-field parameter vectors, ${\bf p}_{li,l'j} = \langle  \hat{\bf p}_{li,l'j}\rangle$ and  ${\bf h}_{li,l'j} = \langle  \hat{\bf h}_{li,l'j}\rangle$ respectively, and where matrix $M = C^a$ or $T$ depending on the nature of the coupling in the intralayer or the interlayer terms of the Hamiltonian.

As the parton transformation 
introduces unphysical states not present in the original spin Hamiltonian, 
it is necessary to introduce  a no-double-occupancy constraint in addition to the usual constraint of having only one fermion per site:
\begin{eqnarray}
n_{i\sigma} = f_i^\dagger f_i =1,~~
f_{i \uparrow} f_{i\downarrow} =f^\dagger_{i\uparrow} f^\dagger_{i\downarrow}=0.
\nonumber \\
\end{eqnarray}
Since it is very difficult to enforce these constraints exactly, we treat them at the mean-field level by introducing a set of Lagrange multipliers $\{ \lambda^a\}$ and adding
\begin{eqnarray}
    {\mathcal H}_\lambda &=&   
    \sum_i  \lambda^z_i ( 1-  f^\dagger_{i} f_{i}  ) 
    \nonumber \\
    &+& \sum_i (\lambda^x_i + i \lambda^y_i)  f^\dagger_{i\uparrow} f^\dagger_{i\downarrow} 
    +\sum_i (\lambda^x_i - i \lambda^y_i) f_{i\downarrow} f_{i\uparrow},
\end{eqnarray}
to the mean-field Hamiltonian ${\mathcal H} = {\mathcal H}_K+ {\mathcal H}_J + {\mathcal H}_\lambda$.

Up to now, the mean field theory is not constrained and any type of solution, {\it Ansatz}, is possible, either a magnetically ordered state or a QSL. As discussed in detail in the rest of the paper, we have solved numerically the self-consistent equations in the Fourier space and for various interlayer couplings and number of layers. Anticipating the discussions, we found that at the saddle points of our model, satisfying the no-double-occupancy constraint in average, that all mean field parameters are at least translationally invariant. We can thus simplify our notations by dropping the site indices such as, for the intralayer mean-field parameters one gets ${\bf p}_{li,l',j} \to {\bf p}_{l,\alpha}$ and ${\bf h}_{li,l',j} \to {\bf h}_{l,a}$, since all nearest neighbor bonds $\langle i,j \rangle_a$ carry the bond orientations. Similarly, all the interlayer mean-field parameters are now depending only on the layer indices $l$ and $l'$ such that ${\bf p}_{li,l',j} \to {\bf p}_{l,l'}$ and ${\bf h}_{li,l',j} \to {\bf h}_{l,l'}$. To avoid any confusion in the following, we will precise whether we are considering interlayer or intralayer parameters.

\subsection{Projective symmetry group}

The quadratic part of the mean-field Hamiltonian in terms of Abrikosov fermions satisfies a projective symmetry group (PSG). In order to analyze this PSG for the multilayer Kitaev spin liquids as done in the next section, we first focus on the single layer Kitaev and follow the construction and notations introduced in [\onlinecite{you_doping_2012}]. Let us here briefly recall their main results for the single layer,  omitting the irrelevant layer index $l$ in this case. The Abrikosov decomposition of the spin at site $i$ introduced above in the text as $S_i^a = \frac{1}{2} f_i^\dagger \sigma^a f_i$  with $f_i = ( f_{i \uparrow} , f_{i \downarrow} )$, the spinon operator $f_{i}$ possesses an additional SU(2) gauge structure that is easier to see by arranging the matrix as\cite{hermele_su2_2007}:
\begin{eqnarray}
F_i = \begin{bmatrix} f_{i \uparrow} & - f_{i \downarrow}^\dagger \\ f_{i \downarrow} &f_{i \uparrow}^\dagger \end{bmatrix}.
\end{eqnarray}
In function of this spinon matrix, the spin can be recast as 
\begin{eqnarray}
S_i^a = \frac{1}{4} \text{Tr} F_i^\dagger \sigma^a F_i.
\end{eqnarray}

In general, the quadratic mean-field part of the Kitaev Hamiltonian can be 
expressed in terms of the $F_i$ matrices \cite{choi_topological_2018} as:
\begin{equation}
{\mathcal H}_K = -\frac{K}{2}\sum_{\langle i j \rangle_a} Tr \{ \sigma^\alpha F_i U_{\langle ij \rangle_a}^\alpha F^\dagger_j \}, 
\end{equation}
where a sum over $\alpha=0, x, y, z$ is assumed as well as a sum over the three bonds of the honeycomb lattice. The four link variable matrices, $U_{\langle ij \rangle_a}^\alpha$, encode the singlet and triplet variational parameters:
\begin{eqnarray}
{U}^0_{\langle i,j \rangle_a} &=&
\begin{bmatrix}
h^0_{ij} & -p^{0*}_{ij} \\
-p^0_{ij} &  -h^{0*}_{ij} 
\end{bmatrix}, \\
U^a_{\langle i,j \rangle_a} &=&
\begin{bmatrix}
h^a_{ij} & -p^{a*}_{ij} \\
p^a_{ij} &  h^{a*}_{ij} 
\end{bmatrix},~
U^b_{\langle i,j \rangle_a} = 
\begin{bmatrix}
-h^b_{ij} & p^{b*}_{ij} \\
-p^b_{ij} &  -h^{b*}_{ij} 
\end{bmatrix}, \nonumber
\end{eqnarray}
here $b\ne a$.

The {\it Ansatz} for the KSL can be obtained by using the PSG of the
Kitaev state. Under PSG the spinon matrix should transform as:
\begin{equation}
F_i\rightarrow R^\dagger_g F_{g(i)} G_g(g(i)),
\end{equation}
where $ R_g$ and $G_g$ are the symmetry and gauge transformation
associated to the group symmetry $g$, respectively.  The mean-field Hamiltonian 
remains invariant under the PSG transformation. For example, we have:
\begin{align*}
{\mathcal H}_{K}&\rightarrow - \frac{K}{2} \\
&\times \sum_{\langle i,j \rangle_a} Tr \{ R_g\sigma^\alpha R^\dagger_g F_{g(i)} G_g(g(i)) U_{\langle i,j \rangle_a}^\alpha G^\dagger_g(g(j)) F^\dagger_{g(j)} \} 
\\
&= -\frac{K}{2} \sum_{\langle i,j \rangle_a} Tr \{ \sigma^\alpha F_i U^\alpha_{\langle i,j \rangle_a} F^\dagger_j \}.
\end{align*}

 In a spin liquid, translational invariance is satisfied so that we can simplify the notations by retaining in the bond direction the subscript of the component, {\it i. e.} $ \langle i,j\rangle_a \equiv a$. The four variable matrices $U_{\langle ij \rangle_a}^\alpha \equiv U_{a}^\alpha $ per link are expressed in terms of singlet and triplet variational parameters, respectively ${\bf p}_{\langle ij \rangle_a}^b \equiv {\bf p}_{a}^b$ and ${\bf h}_{\langle ij \rangle_a}^b \equiv {\bf h}_{a}^b$. We keep in mind that subscripts always refer to the bond direction while superscripts to the operator components.
 
Consider the $C_6^2$ transformation of the 
$z$-link variational parameters of the Kitaev model. This PSG operation should lead to the $x$-link variational parameters in agreement with the Kitaev {\it Ansatz} which can be summarized as:
\begin{align*}
{U}^0_{x} &=
\begin{bmatrix}
-h^{0}_{x} & p^{0*}_{x} \\
p^{0}_{x} &  h^{0*}_{x} 
\end{bmatrix}
= -
\begin{bmatrix}
\Im (h^0_z)& \Re (h^0_z) \\
\Re (h^0_z) &  \Im (h^0_z) 
\end{bmatrix},
\\
{U}^x_{x} &=
\begin{bmatrix}
-h_x^x &  p_x^{x*} \\
-p_x^x &  -h_x^{x*}
\end{bmatrix}
= 
- \begin{bmatrix}
\Re (h^z_z) & \Im (h^z_z) \\
\Im (h^z_z) &  \Re (h^z_z)
\end{bmatrix},
\\
{U}^y_{x}&=
\begin{bmatrix}
h_x^y & -p_x^{y*} \\
p_x^y &  h_x^{y*} 
\end{bmatrix}
= 
-i\begin{bmatrix}
\Re (p^x_z) & \Im (p^x_z) \\
-\Im(p^x_z) &  \Re (p^x_z)
\end{bmatrix},
\\
{U}^z_{x} &=
\begin{bmatrix}
h_x^z & -p_x^{z*} \\
p_x^z &  h_x^{z*} 
\end{bmatrix}
= 
-i \begin{bmatrix}
\Re (p^y_z) &\Im (p^y_z) \\
- \Im (p^y_z) &  -\Re (p^y_z)
\end{bmatrix}.
\\
\end{align*}

 These expressions allow to obtain the $x$-link from the
 the $z$-link variational parameters at each Kitaev layer $l$. Similarly, the $y$-link variables can be obtained by applying a $g=C_6^4$ rotation on the $z$-link.
From this analysis one finds that the non-zero mean-field parameters at each Kitaev layer should satisfy the following relations\cite{choi_topological_2018}:
\begin{eqnarray}
h^0_{l,x} &=& h^0_{l,y}=h^0_{l,z} \equiv h^0_l \in \Im,
\nonumber\\
 h^z_{l,z}  &=&  h^0_l,
\nonumber \\
p^x_{l,z} &=& -i p_{l,z}^y \in \Im,
\nonumber \\
 p^x_{l,x} &=& h_{l,z}^z=h^0_l, 
\nonumber \\
h^z_{l,x} &=& -i p_{l,z}^y,
\nonumber \\
p^y_{l,x} &=& i p_{l,z}^x=p_{l,z}^y \in \Re,
\nonumber \\
p^x_{l,y} &=&  -i p_{l,z}^y,
\nonumber \\
p^y_{l,y} &=& i h_{l,z}^z = i h^0_l,
\nonumber \\
h_{l,y}^z &=& p_{l,z}^x=- i p_{l,z}^y.
\nonumber \\
\label{eq:kitaevansatz}
\end{eqnarray}
Note that the Kitaev {\it Ansatz} at each layer can be expressed in terms of 
only two parameters $h^0_l, p_{l,z}^y$. 

In fact, the PSG analysis is much simpler in the Majorana language as detailed in App.~\ref{app:psg} for the single and two  layer systems. Here, we want to give a quick view about how it works and connects with the present Abrikosov theory. Hence, we drop the layer index $l$ for the discussion below until the end of this section.
The Kitaev model is solved exactly\cite{kitaev_anyons_2006} by introducing four Majorana fermions $\chi_i^\alpha$ ($\alpha = 0,x,y,z$) on each site of the lattice and by mapping the spin operators as $S_i^a = i \chi_i^0 \chi_i^a$. The Majorana operators act on a 4-dimensional Fock space whereas the Hilbert space of the spin is identified with a 2-dimensional subspace of this space. To project onto the physical subspace, the state has to obey the physical constraint $\chi_i^0\chi_i^x\chi_i^y\chi_i^z = 1/4$\cite{kitaev_anyons_2006}. Depending on the $SU(2)$ gauge, those Majorana fermions have been shown to be related to the Abrikosov fermions $f_{i \sigma}$ by the simple relation\cite{burnell_su2_2011}
\begin{eqnarray}
F_i = \frac{1}{\sqrt{2}} ( \chi_i^0 \sigma^0 + i  \chi_i^x \sigma^x + i \chi_i^y \sigma^y +i  \chi_i^z \sigma^z )
\end{eqnarray}
corresponding to specific Majorana modes of the Abrikosov fermions $f_{i \uparrow} = ( \chi_i^0  + i \chi_i^z)/\sqrt{2}$ and $f_{i \downarrow} = ( i \chi_i^x  - \chi_i^y)/\sqrt{2}$. In terms of Majorana bilinears, the Hamiltonian reads
\begin{eqnarray}
{\mathcal H}_K= -K \sum_{a} \left[ i u^a_a \chi_i^0 \chi_j^0 + i u^0_a \chi_i^a \chi_j^a  -  u^0_a u^a_a \right],
\end{eqnarray}
with the $a$-link amplitude: $ u^\alpha_{a} = \langle \frac{i}{2}  \chi_i^\alpha \chi_j^\alpha  \rangle$ ($\alpha=0,x,y,z$)  is a mean field {\it Ansatz} at the $\langle i j \rangle_a$ link.
These are obtained self-consistently giving: $u^0_{ij} =  -0.262433$, 
$u^\alpha_{ij} =  1/2$ if $\alpha =a$ and 0 otherwise. Note that changing the signs of all  bond parameters yields a second possible solution. 

The connection between the Abrikosov and Majorana representation is straightforward with the relations:
\begin{eqnarray}
u_a^0 &=&  - i h^0_z -  p_z^y, \nonumber \\
u_a^a &=&  - i h^0_z + p_z^y,
\end{eqnarray}
being readily satisfied. Thus, the the $x$ and $y$-bond amplitudes can be expressed in terms of just $h_z^0$ and $p_z^y$ on the $z$-link.  The rest of the amplitudes of the lattice are straightforwardly obtained using translation symmetries (see App.~\ref{app:psg}). This procedure can also be extended to the multilayer case as we have done for the PSG on the bilayer system, where it is shown that the ground state {\it Ansatz} consists in two copies of the single layer but with opposite signs on all bond parameters. Interestingly, this solution requires a specific interlayer inversion symmetry that is not the expected simple reflection symmetry. Solving the self-consistent equations for the bilayer case, we have indeed checked that it is the lowest stable solution reachable.

\section{Quantum spin liquid solutions in multilayered Kitaev models}
\label{sec:qsl}

We investigate possible quantum spin liquid states arising in multilayered Kitaev models. 
For this purpose we solve ${\mathcal H} = {\mathcal H}_K +{\mathcal H}_J$, self-consistently neglecting the magnetic channel. We analyze the total mean-field energy per unit cell: $E=  {\mathcal H}/N $ by solving self consistently the mean field equations of the system self-consistently, and the magnitude of the total spin at each site of an effective honeycomb model (see Fig.~1 and caption):
\begin{equation}
S_{tot}^2=\left( \sum_{l=1}^n {\bf S}_{l} \right)^2.
\end{equation} 

From the dependence of the mean-field parameters on $J$ we identify various phase transitions. We obtain the excitation spectrum of the system which allows concluding on the character of the multilayered Kitaev model, in particular, whether the system is gapped or not. We consider FM ($J>0$) and AFM ($J<0$) coupled multilayered Kitaev models. From now on we assume $K \equiv 1$ unless otherwise stated. 

\subsection{Projective symmetry group}

For the interlayer contributions, the Heisenberg terms read:
\begin{equation}
{\mathcal H}_J= -\frac{J}{2} \sum_{i,l,l'} Tr \{ \sigma^b F_{l,i} V_{ li,l'i}^b F^\dagger_{l',i} \}, 
\end{equation}
with $l'=l+1$ as in the interlayer interaction in (\ref{eq:model}). The $V$-matrices are now:
\begin{align*}
V^{0}_{ li,l'i } =
\begin{bmatrix}
h^{0}_{li,l'i} & -p^{0*}_{li,l'i} \\
-p^{0}_{li,l'i} &  -h^{0*}_{li,l'i}
\end{bmatrix},~
V^{a}_{li,l'i} =
\begin{bmatrix}
h_{li,l'i}^{a} & -p_{li,l'i}^{a*} \\
-p^{a}_{li,l'i} &  h_{li,l'i}^{a*} 
\end{bmatrix}.
\end{align*}

Our self-consistent solutions consist on Kitaev-type QSLs at each layer displaying flux patterns corresponding to the single layer Kitaev model whose signs alternate from layer to layer. This flux sign alternation is in agreement with previous analysis of the AF coupled Kitaev bilayer \cite{seifert_bilayer_2018}. 
Our analysis finds that such flux sign alternation can be extended to an arbitrary number of layers. This means that the variational parameters of each Kitaev layer is related with the consecutive layers as:   
\begin{eqnarray}
h^0_{l+1} &=& (-1)^{l+1} h^0_l, 
\nonumber \\
h_{l+1,x}^z &=& h_{l+1,y}^z= i (-1)^{l} p_{l,z}^y,
\nonumber \\
h_{l+1,z}^z &=& (-1)^{l+1} h^0_l,
\nonumber \\
p_{l+1,x}^x &=& (-1)^{l+1} h^0_l,
\nonumber \\
p_{l+1,x}^y &=& (-1)^{l+1} p_{l,z}^y,
\nonumber \\
p_{l+1,y}^x &=& i(-1)^{l} p_{l,z}^y,
\nonumber \\
p_{l+1,y}^y &=& i(-1)^{l} h^0_l, 
\nonumber \\
p_{l+1,z}^x &=& i(-1)^{l} p_{l,z}^y, 
\nonumber \\
p_{l+1,z}^y &=& (-1)^{l+1} p_{l,z}^y.
\label{eq:ansatz}
\end{eqnarray}

Among all possible interlayer variational parameters 
we choose an {\it Ansatz} leading to numerically stable self-consistent solutions. For $J>0$ only the triplet pairings $p^{a}_{l,l+1} \ne 0 \in \Re$ are non-zero while 
for $J<0$ we take $h^0_{l,l+1} \ne 0 \in \Im$. 
These are not the only possible {\it Ans\"atze} for the 
interlayer parameters, but other choices can be 
related to these through gauge
transformations\cite{lee_doping_2006}.

\subsection{Ferromagnetic interlayer coupling}

\begin{figure}[h!]
    \centering
\includegraphics[width=8.5cm,clip]{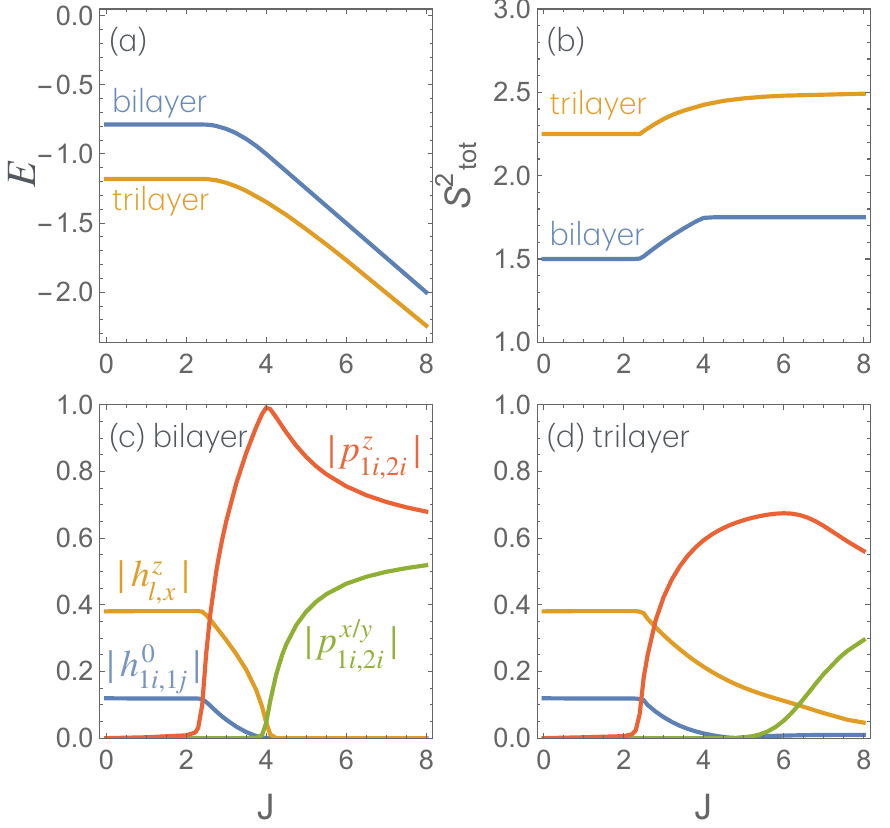}
    \caption{Dependence of the ground state properties of Kitaev multilayers on the FM interlayer coupling $J$. In (a) the dependence of the ground state energy in Kitaev
    bilayers (blue solid line) and trilayers (red solid line) on $J$ is shown. In (b) the dependence of the total spin magnitude, $S_{tot}^2=(\sum_l {\bf S}_{l,i})^2$, on $J$ is plotted. (c) and (d) show the dependence of the mean-field parameters 
    in the Kitaev bilayer and trilayer, respectively. $K=1$ in these plots.
    }
\label{fig:qsl_fm}
\end{figure}

In Fig.~\ref{fig:qsl_fm} (a), (b) we show the dependence of  $E$ and $S_{tot}^2$ on 
the interlayer FM coupling, $J > 0$, in bilayer and trilayer Kitaev models. A direct inspection of 
the plots shows that bilayer and trilayer systems behave quite differently. While in 
Kitaev bilayers, 
$S_{tot}^2$ displays two separate transitions, in Kitaev trilayers, a single transition occurs.
The ground state energy is independent of $J$ up to $J\approx 2.5$ in both bilayer and trilayer Kitaev models. In this regime, $S_{tot}^2=n S (S+1)={3 n \over 4}$, which means that the Kitaev layers are effectively decoupled and the ground state is the
direct product of the Kitaev spin liquids at each layer. In the large-$J$ limit of a bilayer system, the neighbor spins of 
the two adjacent planes lock in a triplet state so that a crystal of weakly coupled interlayer triplets is 
formed. In contrast, trilayer Kitaev models, host a QSL with effective spin,
$S_{tot} > {3 n \over 4}$ no matter the value of $J$ is. Note that $S_{tot}^2 \rightarrow 2$ in a pure triplet state 
in a bilayer in contrast to $S_{tot}^2 \rightarrow 1.75 $ as $J \rightarrow \infty$ shown in Fig.~\ref{fig:qsl_fm} (b). Similarly, Kitaev trilayers at $J \rightarrow \infty $, lead to $ S_{tot}^2 ={5 \over 2}$ which is much smaller than the exact $S_{tot}^2 ={15 \over 4}$ of an isolated trimer as shown in Fig.~\ref{fig:qsl_fm} (b). 
The departure of $S_{tot}^2$ in the $J \rightarrow \infty$ limit from the exact value for isolated dimers or trimers is due to the mean-field treatment of the no double occupancy constraint implicit in the approach used. 

The different behavior of Kitaev bilayers and trilayers is
actually generic of even vs. odd layered Kitaev models and can be further analyzed by inspecting the 
dependence of the self-consistent 
mean-field parameters on $J$.
In Fig.~\ref{fig:qsl_fm} (c), (d) the dependence of the non-zero variational parameters with the interlayer coupling $J$ of bilayer and trilayer Kitaev models is compared. In the bilayer case the interlayer coupling is zero up to a critical $J_{c1} =2 $ at which the interlayer triplet pairing parameter,  $p^{z}_{1i,2i}$, increases and the in-layer parameters are suppressed until they are zero at $J_{c2}$. For $J>J_{c2} \approx 4$  the interlayer triplet pairings,  $p^{x/y}_{1i,2i}$, also become non-zero. This indicates that there are three different phases as $J$ increases in Kitaev bilayers. In contrast, in Kitaev trilayers there is no phase at large $J$ at which the intralayer variational parameters are all zero implying that a QSL survives for any $J$. 

\begin{figure}[t!]
    \centering
  \includegraphics[width=8.8cm,clip]{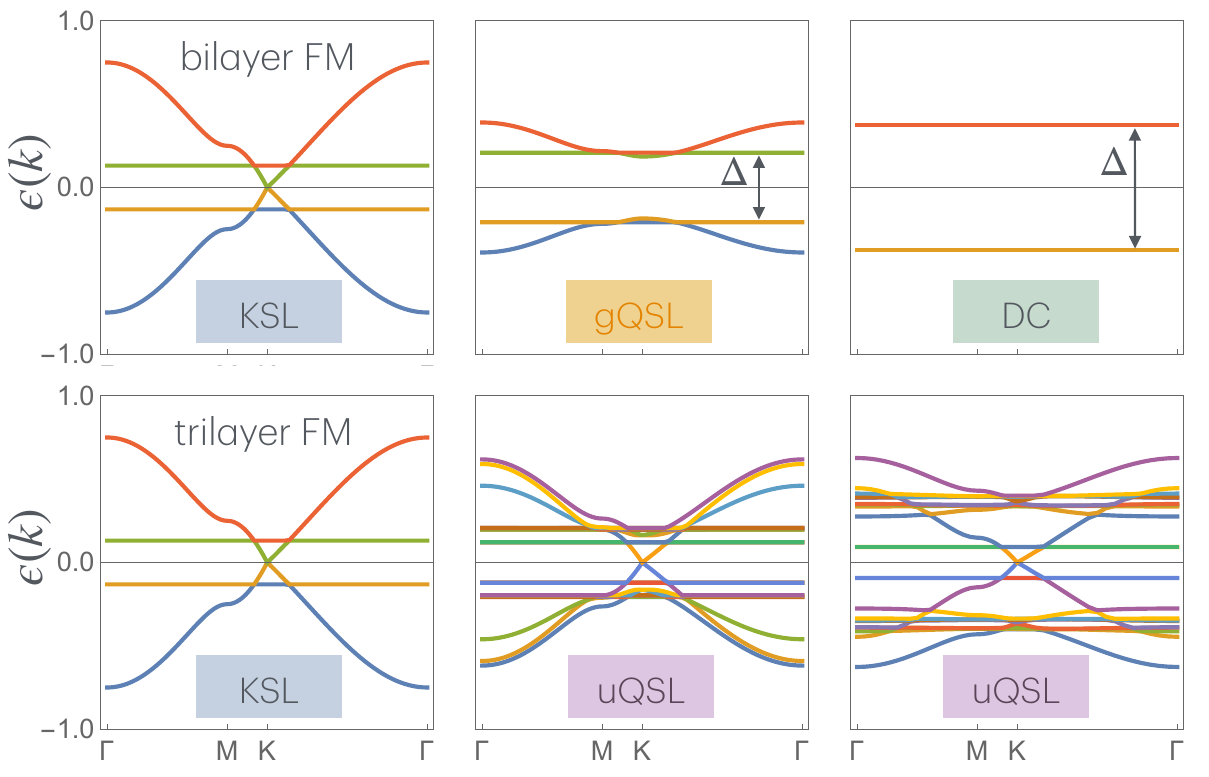} 
    \caption{Band structure of a Kitaev FM bilayer (top) and trilayer (bottom) as a function of $J$. For the bilayer, a gap $\Delta$  opens and a gapped QSL (gQSL) emerges for $ J > J_{c_1} \approx 2.5$. For $J \geq J_{c_2} \approx 4$, a phase (DC), with ferromagnetic valence bonds form between sites on two different layers. For the trilayer,  there is no gap opening and an ungapped quantum spin liquid (uQSL) is stable for any $J>J_{c_1}$. While the gQSL is generic of even-layered Kitaev models the uQSL is generic of 
    odd-layered systems. $K=1$ in these plots.}
\label{fig:bitrilayer}
\end{figure}

In order to gain physical insight on the different phases encountered we have explored the behavior of excitation spectra across the various transitions observed in bilayers and trilayers. The spectra 
of decoupled bilayers for $J<J_{c1}$ shown in Fig.~\ref{fig:bitrilayer} (upper row) is that of the KSL which 
displays dispersing bands with Dirac nodes at $K$-points and flat bands corresponding to localized flux 
excitations in agreement with the Kitaev exact solution, as it should. While the dispersing bands are doubly degenerate 
the flat bands are six-fold degenerate since we have two independent Kitaev models. As $J$ increases, in 
the intermediate regime $J_{c1} < J < J_{c2} $, a gap opens up at the Dirac cones. For $J>J_{c2}$ the 
dispersing bands flatten out and merge with the flux bands so that Majorana fermions become localized describing 
a set of decoupled FM interlayer dimers arranged on a honeycomb lattice. Since
intralayer variational parameters are non-zero, the phase in the $J_{c1} < J < J_{c2}$ can be regarded as a gQSL. For $J>J_{c2}$ the QSL variational parameters become zero indicating the DC phase. 

The spectra of Kitaev trilayer behaves quite differently. As shown in Fig.~\ref{fig:bitrilayer} (lower row) the most important difference with Kitaev bilayer
is that Dirac cones persist even for $J>J_{c1}$, indicating the survival of an uQSL at arbitrarily 
large $J$. Increasing $J$ leads to a splitting of the dispersing bands and the flat bands breaking their 
degeneracy but Dirac cones are always present. This means that the QSL formed in this regime cannot be regarded as a pure KSL.
However, certain features of the single layer KSL solution, such as the Dirac cones, survive in the presence of $J$. 

In summary, Kitaev bilayers host a gapped QSL (gQSL) with $S_{tot}^2 > {3 n \over 4}$, {\it i.e.} 
with effective total spin $S_{tot} >1/2$ which emerges in 
an intermediate range of FM couplings: $J_{c1} < J < J_{c2} $. 
In contrast, Kitaev trilayers only host an uQSL 
which survives all the way up to $J \gg J_{c1}$. The different behavior
of the two systems is reflected in the phase diagram of Fig.~\ref{fig:structure}.

\subsection{Antiferromagnetic interlayer coupling}

\begin{figure}[h!]
\centering
\includegraphics[width=8.5cm,clip]{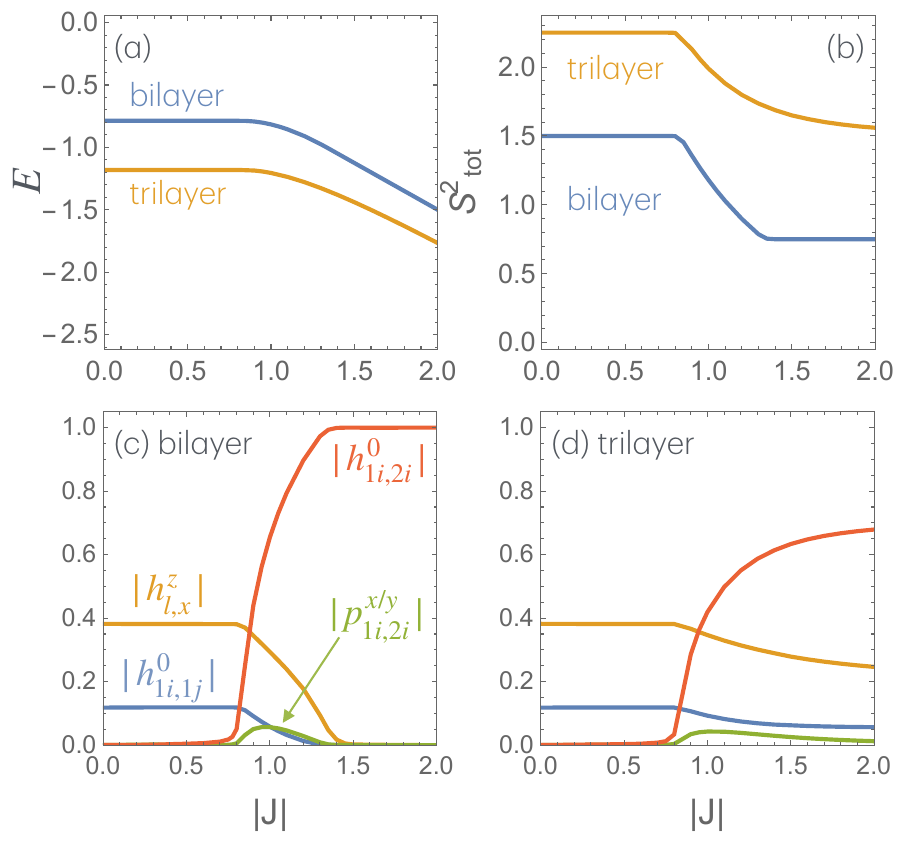}
    \caption{Dependence of the ground state properties of Kitaev multilayers on the AFM interlayer coupling, $J<0$. In (a) the dependence of the ground state energy of Kitaev
    bilayers (blue solid line) and trilayers (red solid line) on $J$ is shown. In (b) we show the dependence of the total spin magnitude, $S_{tot}^2=(\sum_l {\bf S}_{l,i})^2$, on $|J|$.  (c) and (d) show the dependence of the mean-field parameters in bilayers and trilayers, respectively. $K=1$ in these plots.
    }
\label{fig:qsl_afm}
\end{figure}

We now consider AFM coupled, $J<0$, Kitaev multilayers. 
In Fig.~\ref{fig:qsl_afm} (a), (b) we show the dependence of the mean-field energy, $E$, and 
$S_{tot}^2$ on the absolute value of the interlayer AFM coupling, $|J|$, of 
bilayer and trilayer Kitaev models. Similarly to FM coupled multilayers, AFM coupled bilayer 
and trilayer Kitaev models behave quite differently. While in Kitaev bilayers, 
$S_{tot}^2$, displays two separate transitions, in Kitaev trilayers, a single transition occurs.
The ground state energy is independent of $J$ 
up to $J =J_{c1} \approx 0.75$ in 
bilayer and trilayer Kitaev models. In this regime, 
$S_{tot}^2=n S (S+1)={3 n \over 4}$, so the 
Kitaev layers are effectively 
decoupled and the ground state is the
direct product of the Kitaev spin liquids on 
each layer. In the large-$J$ 
limit of bilayer models, the neighbor spins of 
the two adjacent planes lock in a singlet state so that a decoupled 
crystal (DC) of interlayer singlets is formed. In contrast, trilayer AFM 
coupled Kitaev multilayers, host a QSL with effective spin,
$S_{tot}  < {3 n \over 4}$ for $J > J_{c1}$. Note that 
$S_{tot}^2 \rightarrow 0$ for an isolated singlet state formed between two adjacent spins in contrast to 
$S_{tot}^2 \rightarrow {3 \over 4} $ shown in Fig.~\ref{fig:qsl_afm} (b). Similarly,
a larger value than the exact $S_{tot}^2 ={3 \over 4}$ expected is 
found in strongly AFM coupled trilayers instead of $S_{tot}^2 ={3 \over 2}$ 
shown in Fig.~\ref{fig:qsl_afm} (b). As discussed in FM coupled Kitaev models, the departure of $S_{tot}^2$ from the exact value expected for an isolated singlet is due to the mean-field 
treatment of the no double occupancy constraint. Further discussion about this point 
is provided in the next section.

The dependence of the self-consistent mean-field parameters 
on $|J|$ is shown in Fig.~\ref{fig:qsl_afm} (c), (d). The results are qualitatively similar to Kitaev multilayers coupled through a FM interlayer coupling of Fig.~\ref{fig:qsl_fm} (c), (d). The AFM coupled Kitaev bilayers host an intermediate gQSL in the range, $J_{c1} < J < J_{c2}$, between KSL and DC phases as in the FM, $J>0$, case. In contrast, in Kitaev trilayers a single transition from the KSL to the uQSL occurs at $J > J_{c1}$.

\subsection{Multilayers}

We have increased the number of layers up to five layers finding that the even-odd effect found for bilayer vs. trilayer systems is generic of arbitrary number of Kitaev layers. This can be understood from the discussion in the previous section. In the even layered case, all the spins in consecutive layers can dimerize either 
into singlets or triplets depending on the sign of $J$. Since all spins are involved in the formation of interlayer dimers a gap can open up in the spectra at sufficiently large $J$. In contrast, the odd layered system effectively behaves as a single Kitaev layer since not all spins can simultaneously form interlayer dimers even at large $J$.
Thus, while even-layered Kitaev model hosts a DC at large $J$, 
odd-layered Kitaev multilayer hosts an ungapped QSL which survives 
to arbitrarily large $J$ values. An intermediate gQSL phase emerges between 
the KSL and the DC in the even-layered case. 
The phase diagrams of Fig.~\ref{fig:structure} summarize all our main results
which are common to both FM and AFM coupled Kitaev multilayers. However, there 
are quantitative differences in the
critical values of the FM and AFM cases. While in FM coupled bilayers,
$J_{c1}=2, J_{c2}=4$ in AFM  coupled bilayers $J_{c1}=-0.75, J_{c2}=-1.45$.
The $J_{c1} =2$ in FM trilayer systems is larger than $J_{c1}=-0.75$ in the AFM case.

%\subsection{Decoupled Kitaev magnets}

%\subsection{Even-layered Kitaev magnets}

%\subsection{Odd-layered Kitaev magnets}

%\section{Effect of an external magnetic field}

%As shown by Kitaev when a weak external uniform magnetic field in the $(1,1,1)$ direction is applied to the Kitaev model a gap opens up leading to a topological state characterized by a Chern number $\nu= \pm 1$. Here we 
%explore the effect of such magnetic field on the spectra of multilayered Kitaev models. 

\section{Gutzwiller projection and the adiabatic continuation to the exact ground state}
\label{sec:interlayer}

In order to test the reliability of the parton approach 
we compare our Abrikosov mean-field ground state properties with 
ED on small clusters. The ground state mean-field energies are 
found to be quite above the exact energies due to 
the double occupancy constraint which is treated at the
mean-field level. However, by removing 
doubly occupied configurations from the mean-field QSL wavefunctions at
large $J$ we find how that their energies are greatly improved 
becoming very close to the exact energies. Indeed, such 
Gutzwiller projected energy of the multilayer Kitaev model interpolates 
between the exact energy of uncoupled Kitaev layers and the 
DC at large-$J$.
The comparison between mean-field and ED energies can 
be found in Fig.~\ref{fig:energybilayers} which are 
discussed in a case by case basis below.
Details on the calculations related to the results shown
in this section can be found in App. \ref{app:largeJ}.

\subsection{Bilayers}

{\it Ferromagnetic ($J>0$) interlayer coupling.} 

In this case, in the DC phase ($K=0$), the bilayer system becomes a set of $2N$ 
decoupled dimers consisting of triplets formed between the spins in adjacent sites on the two layers. Hence, the exact ground state is just the product state of these
interlayer triplets: ${| \Psi_0 \rangle =  \Pi_{i,S^z} |\psi^{S^z}_{i} } \rangle$ where $S^z$ identifies 
each of the triplet states $S^z=0, \pm 1$ at a give site: 
\begin{eqnarray}
|\psi^{0} \rangle = {1 \over \sqrt{2} } (| \uparrow_1  \downarrow_2 \rangle +
| \downarrow_1 \uparrow_2 \rangle ),
\nonumber \\
| \psi^{+1} \rangle = |\uparrow_1\uparrow_2 \rangle,
~~
| \psi^{-1} \rangle = |\downarrow_1 \downarrow_2 \rangle,
\end{eqnarray}
written in the basis $| \sigma_l \sigma'_{l'} \rangle$ with $l$ and $l'$ the layer indices and  where we have omitted the $i$-site label for simplicity, since we 
assume uniform triplet formation between the two layers.

\begin{figure}[t!]
    \centering
 \includegraphics[width=8.5cm,clip]{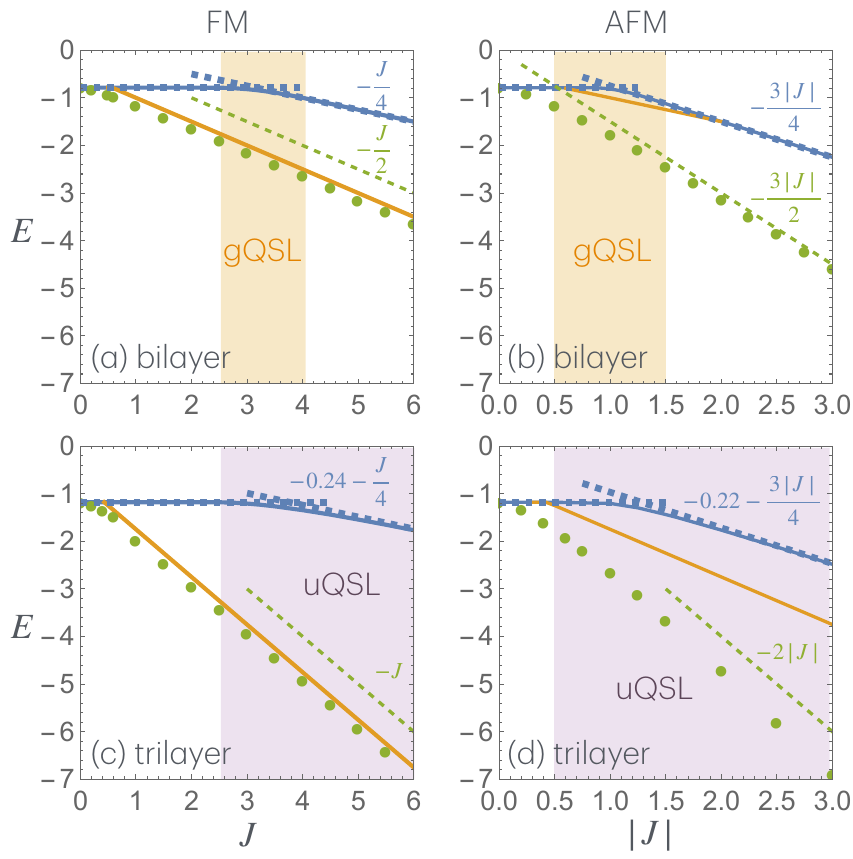} 
    \caption{Dependence of ground state energies of the bilayer (a,b) and the trilayer (c,d) Kitaev models with interlayer FM (left column) and AFM (right column) coupling $J$.
    The Abrikosov mean-field QSL energy (blue circles)  is compared with the energy with magnetic order  
    (yellow lines) and the exact ED energy (green circles).
     (a) and (b): for FM  coupling, $J>0$, the energy of the FM mean-field solution follows closely the ED energy. (c) and (d): for AFM coupling, $J<0$, the energy of the mean-field QSL and AFM solutions are quite above the ED energy. The dashed green lines indicate the energy dependence in the large $J$ limit. $K=1$ in these plots.} 
\label{fig:energybilayers}
\end{figure}

Hence, the exact ground state of the bilayer is $3^{2N}$-fold degenerate 
for $J>0$ and $K=0$. Thus, the total energy per unit cell reads:
\begin{equation}
{E} = -{ J \over 2}.
\end{equation}
with $J >0$.
Since at $J=0$, the system becomes a set of decoupled 
Kitaev spin liquids, we expect that the energy 
of the system interpolates between 
$E = -0.787$ at $J=0$ and ${E} = -{J \over 2}$ at $K=0$. 
However, Fig.~\ref{fig:energybilayers} (a) shows how the exact ground state energy converges to $E \sim -0.65 - {J \over 2}$ rather than $-{J \over 2}$ when $J > (0.4-0.5)$. 
This indicates that the triplets are not completely decoupled but
a remnant Kitaev interaction energy between the effective
$S=1$ triplets at each lattice site of the honeycomb lattice 
survives. Indeed, such extra $-0.65$ energy contribution 
has been previously found in the $S=1$ Kitaev model \cite{tomishige_interlayer_2018}.
At $K=0$ the ED recovers the expected $-{J\over 2}$ triplet energy
of uncoupled triplets, as it should.

Although the mean-field energies coincide with the exact energies 
of decoupled Kitaev layers, at large $J$, the mean-field energy behaves as 
$-J/4$ which is half of the exact ED energy, $-J/2$, of isolated triplets (see Fig.~\ref{fig:energybilayers} (a)). The mean-field ground state wavefunction at $K=0$ contains doubly-occupied configurations:
\begin{equation}
   |\Psi^{MF}_0 \rangle = {1 \over 2} (f^\dagger_{1 \uparrow}  f^\dagger_{1 \downarrow} - f^\dagger_{1 \uparrow}  f^\dagger_{2 \downarrow} - f^\dagger_{1 \downarrow}  f^\dagger_{2 \uparrow}  - f^\dagger_{2 \uparrow}  f^\dagger_{2 \downarrow} ) |0 \rangle,
\end{equation}
since it only has to satisfy the no-double occupancy constraint on average: $\langle n_{l\sigma} \rangle= {1 \over 2}$ , $\langle \Psi^{MF}_0 |f^\dagger_{1 \uparrow } f^\dagger_{1 \uparrow} | \Psi^{MF}_0 \rangle =0 $. Note how the Gutzwiller projected ground state: $| \Psi_G \rangle =\Pi_i n_{i} (2-n_{i} ) |\Psi^{Mƒ}_0 \rangle $ with $n_i= n_{i\uparrow} + n_{i\downarrow}$, coincides with the exact, $S=1, S^z=0$ triplet state
with corresponding $-{J \over 2}$ energy. The two other, $S=1, S^z= \pm 1$, triplet states may be obtained in a similar manner. 

Allowing for the magnetic channel in the mean-field Abrikosov fermion 
approach leads to solutions which reproduce rather closely the ED 
dependence of the energy with $J$ as shown in Fig.~\ref{fig:energybilayers} (a). The magnetic channel leads to solutions in which the two spins in the two layers align ferromagnetically. In principle, one would  naively expect a mean-field energy of $-J/2$ at large $J$. Strikingly, there is an additional contribution which lowers the mean-field energy even further towards the ED energy. 
The origin of this energy comes precisely from a
shift in the bands due to the Lagrange multipliers imposing the constraints. The ferromagnetic state with the two spins aligned is one of the triplet states of the exact solution. 
Hence, allowing for this kind of magnetic order inevitably leads the mean-field solutions closer to the exact energy. In some sense it is similar to having performed the Gutzwiller projection. In doing so there is an energy gain associated with the constraint.  Or in other words, the static magnetic solution mimics, at the mean-field level, more precisely the dynamical gauge fluctuations of the constraint than the quantum spin liquid solution. On the other hand, in the limit of large $J$ we should expect decoupled triplets with no real magnetic order.  Hence, while we have a better estimate of the ground state energy we should keep in mind that the ground state should converge to decoupled triplets at large $J$ rather than to a broken symmetry FM  mean-field solution. Such triplets could be described 
through the Gutzwiller projection of the mean-field QSL solution.  In spite of this, the magnetic solution provides a helpful understanding of the exact solution containing a non-trivial energy contribution from the constraint.  

{\it Antiferromagnetic ($J<0$) interlayer coupling.}

In this case, at $K=0$ the system hosts a set of $2 N$ decoupled interlayer dimers 
locked-in non-degenerate singlets. Hence, the ground state of the system is a product state of non-degenerate 
singlets: ${| \Psi_0 \rangle = \Pi_{i} | \psi^{0}_{i} \rangle }$, with:
$|\psi_i^0 \rangle = {1 \over \sqrt{2} } (|\uparrow_1 \downarrow_2 \rangle +
| \downarrow_1 \uparrow_2 \rangle )$. The ground state energy is, in this case: 
\begin{equation}
E = -{3 |J| \over 2}
\end{equation}
where $J<0$.
As in the FM case, we compare the ED with the mean-field Abrikosov fermion energies in Fig.~\ref{fig:energybilayers} (b). Again in the large $J$ limit, the mean-field energy 
of $-3 J/4$ is half of the exact value associated with decoupled singlets. In 
contrast, the mean-field ground state wavefunction for $K=0$ reads:
\begin{equation}
   |\Psi^{MF}_0 \rangle = {1 \over 2} (f^\dagger_{1 \uparrow}  f^\dagger_{1 \downarrow} + f^\dagger_{1 \uparrow}  f^\dagger_{2 \downarrow} - f^\dagger_{1 \downarrow}  f^\dagger_{2 \uparrow}  + f^\dagger_{2 \uparrow}  f^\dagger_{2 \downarrow} ) |0 \rangle.
\end{equation}
After Gutzwiller projection $|\Psi_G \rangle = P_G |\Psi^{MF}_0 \rangle $ one finds the expected singlet combination between spins on two adjacent layers with energy ${ 3 J \over 4}$. (or ${3 J \over 2}$ per unit cell). Allowing for the magnetic channel in the mean-field Abrikosov approach does not lead to the correct energy as found on FM bilayers. In the AFM case, 
the $| \uparrow_1 \downarrow_2 \rangle$ or 
$|  \downarrow_1 \uparrow_2 \rangle$ configurations
have very different energy from the true singlet state since the latter contains quantum fluctuations. This is in contrast to FM bilayers in which the $| \uparrow_1 \uparrow_2 \rangle$ or $| \downarrow_1 \downarrow_2 \rangle$ configurations have the same energy as the true ground state of the triplet between two spins.

\subsection{Trilayers}

For $K=0$ the FM trilayers become a set of $2 N$ decoupled trimers connecting the three layers. The exact ground state of the whole system, in this case, is a product state of the interlayer trimer ground states 
which is four-fold degenerate. The corresponding four exact ground states formed at each trimer read:

\begin{eqnarray}
    |\Psi_0^{S^z=+{3 \over 2} } \rangle &=& | \ua_1 \ua_2 \ua_3 \rangle
\nonumber \\
 |\Psi_0^{S^z=-{3 \over 2} } \rangle &=& | \da_1 \da_2 \da_3 \rangle
\nonumber \\  
    |\Psi_0^{S^z=+{1 \over 2} } \rangle &=& {1 \over \sqrt{6}} ( | \ua_1 \ua_2 \da_3 \rangle + 2 |\ua_1 \da_2 \ua_3 \rangle + | \da_1 \ua_2 \ua_3 \rangle ),
\nonumber \\
    |\Psi_0^{S^z=-{1 \over 2} } \rangle &=& {1 \over \sqrt{6}} ( | \da_1 \da_2 \ua_3 \rangle + 2 | \da_1 \ua_2 \da_3 \rangle + | \ua_1 \da_2 \da_3\rangle ),
  \nonumber \\  
\end{eqnarray}
which is now four-fold degenerate describing an effective $S=3/2$ 
at each site of a honeycomb lattice. 

The total exact energy per unit cell of the Kitaev 
trilayers at $K=0$ is now: 
\begin{equation}
E = -J.
\end{equation}
Hence, the energy is doubled with respect to FM bilayers due to the energy gain of the triplet which can now resonate between the two bonds available among the three layers. 
The ED energy calculation of Fig.~\ref{fig:energybilayers} (c) shows how the energy at large $J$ rather behaves as: $ {E \over N}  \sim -J-0.9563$. Again a remnant energy contribution from the Kitaev interactions between the effective $S={3 \over 2}$ on the honeycomb lattice is present.

In Fig.~\ref{fig:energybilayers} (c) the mean-field energy of the QSL Abrikosov mean-field solution is compared with ED results. At large $J$ the MF QSL energy behaves as $ \sim -0.24 - {J \over 4} $ in contrast to the 
$-{J \over 4}$ dependence found in FM bilayers. The additional energy gain is due to the 
existence of Kitaev-type QSL behavior on each layer which survives at large-$J$ (this can be 
seen in dispersions of Fig.~\ref{fig:qsl_fm} which display  characteristic Dirac cones). 
Again the QSL mean-field energies depart from the ED energies. The Gutzwiller projection of the mean-field Abrikosov solution would recover the right behavior of the exact energy at large $J$. 

As found in FM bilayers, allowing for magnetism in FM trilayers leads to interlayer FM order as $J$ is increased and a mean-field ground state energy in very good agreement with the ED. This is because the mean-field FM state is just the all up, $| \ua_1 \ua_2 \ua_3  \rangle $, or all down, $| \da_1 \da_2 \da_3  \rangle $ , configurations which are part of the ground state quartet associated with $S=3/2$ interlayer spins. Hence, roughly speaking, allowing for magnetic order in the mean-field Abrikosov theory 
mimics the effect of the Gutzwiller projection. As shown in Fig.~\ref{fig:energybilayers} (c) the FM solution favored at the mean-field level behaves as ${E \over N} =-0.75-J$, quite close to the ED energy. The downward energy shift from $-J$ is associated with the Lagrange multipliers enforcing the constraints.
Hence, a similar discussion to the one given 
above for FM bilayers applies here.  

{\it Antiferromagnetic ($J<0$) interlayer coupling.}

For $K=0$, AFM trilayers behave as a set of $2N$ decoupled trimers whose ground state is doubly-degenerate in contrast to the non-degenerate singlets found in bilayers. The two degenerate ground states at each trimer read:

\begin{eqnarray*}
    |\Psi_0^{S^z=+{1\over 2}}  \rangle = | \ua_1 \ua_2 \da_3 \rangle - 2 |\ua_1 \da_2 \ua_3 \rangle + | \da_1 \ua_2 \ua_3 \rangle , 
\nonumber \\
   | \Psi_0^{S^z=-{1\over 2}} \rangle = | \da_1 \da_2 \ua_3 \rangle - 2 | \da_1 \ua_2 \da_3 \rangle + | \ua_1 \da_2 \da_3\rangle, 
\end{eqnarray*}

describing an effective interlayer $S=1/2$ state interacting through Kitaev interactions within the honeycomb layers. The corresponding total energy per unit cell of AFM trilayers,
at $K=0$, is:
\begin{equation}
E = - 2|J|. 
\end{equation}
The ED energy of AFM trilayers shown in Fig.~\ref{fig:energybilayers} (d) displays a downward shift at large $J$ with respect to the 
$E  = - 2 |J| $ energy of decoupled trimers. Indeed, the ED energy dependence reads: 
${E \over N} = -2 |J|- 0.931$. Such energy lowering 
is of similar magnitude to the one found in FM trilayers and can be attributed to the Kitaev interaction between spins at each honeycomb layer.  

In Fig.~\ref{fig:energybilayers} (d) the Abrikosov mean-field QSL energies are compared with ED energies. At large $J$ the energy behaves as 
$ \sim - 0.22 -{3|J|\over 4} $ in contrast to the 
$-{3 |J| \over 4}$ dependence found in FM bilayers. The additional energy is due to the 
existence of Kitaev type QSL correlations in each layer at the mean-field level which survive at large-$J$ (this can be 
seen in dispersions of Fig.~\ref{fig:bitrilayer}
displaying the characteristic Dirac cones of the KSL). 
Again the Abrikosov mean-field energy of the QSL solution as $|J|$ increases is very different from the ED energy. For large-$|J|$ the Gutzwiller projection of our mean-field {\it Ansatz} would lead to the exact wavefunction for the three spins in the three layers forming effective $S=1/2$ doublets recovering the exact energy. 

Allowing for magnetic order leads to a lower QSL mean-field energy (see Fig.~\ref{fig:energybilayers} (d)). Unlike the FM case, the energy of the AFM solution is still way above the ED energy. This is because the broken symmetry AFM configuration $|\ua_1 \da_2 \ua_3 \rangle$ or $| \da_1 \ua_2 \da_3 \rangle $ picked up by the mean-field calculation is very different from the interlayer exact ground doublet state.
The Gutzwiller projected mean-field state at $K=0$ leads to energies 
converging the ED energies at large $J$ as shown in Fig.~\ref{fig:energybilayers} (d).

\section{Magnetic field effect on multilayer Kitaev models}
\label{sec:magnetic}

We finally explore the effect of a magnetic field on multilayered Kitaev systems. We focus
in the case in which the magnetic field is applied along the $[111]$-direction, ${\bf B} = {(1,1,1) \over \sqrt{3}} B$ since this is the most relevant situation in the single Kitaev layer case \cite{kitaev_anyons_2006}. Apart from the mean-field parameters we also monitor the magnetization: $M=\sqrt{S^{x2}+ S^{y2} + S^{z2}}$ and the gap in the spectrum for different $J$ and $B$. 

\begin{figure}[t!]
\centering
\includegraphics[width=8.5cm,clip]{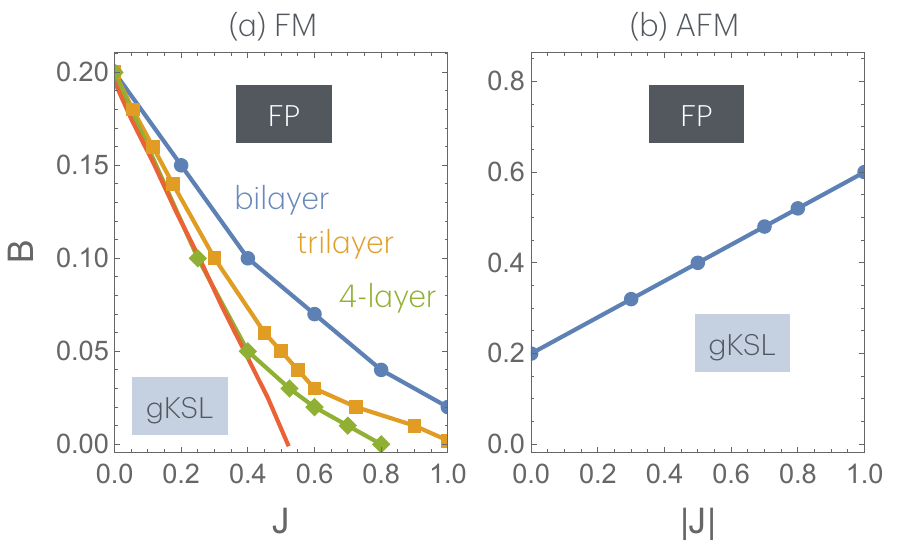} 
    \caption{The $B$ vs. $J$ phase diagram of the (a) FM and (b) AFM multilayered Kitaev models under a $[111]$ magnetic field. (a) A transition from the topological gKSL to the fully polarized state occurs at different transition lines for bilayer (blue), trilayer (yellow), and four layer (green) Kitaev models. The transition from the FP to the gKSL occurs at the same transition line (red) independently of the number of Kitaev layers. (b) Only the bilayer is represented for the AFM coupling. The strong competition between quasi-degenerate 
    ground states does not allow finding sufficiently converged solutions for $J<0$ and $n>2$.} 
\label{fig:phasedfm}
\end{figure}

The effect of $B$ and FM coupled, $J>0$, Kitaev multilayers is summarized  
in Fig.~\ref{fig:phasedfm} for $J>0$ (a) displays a transition between the gKSL induced by a finite magnetic field and the FP state stabilized by a large $B$. A hysteresis region emerges between the topological gKSL and the fully polarized (FP) state. Such region is defined by the solid lines which depend on the  number of layers corresponding to 
$n=2,3,4$ layers and the dotted line which is independent of $n$. It is worth noting that our phase diagram for bilayers is in agreement with previous
Majorana mean-field calculations on bilayers \cite{vijayvargia_topological_2024}.   

In contrast, the effect of a magnetic field on AFM coupled, $J<0$, Kitaev multilayers is quite different. The phase diagram of AF coupled bilayers of Fig.~\ref{fig:phasedfm} (b) shows how the $B$ vs. $J$ transition line between the gKSL and the FP displays a positive instead of a negative slope. This is because a larger magnetic field is needed to fully polarize the system as the AF $J$ increases since it acts oppositely to ${\bf B}$. Thus, the gKSL phase in Kitaev bilayers is stabilized by the AF $J$ in a broader parameter region than by a FM $J$. 

\section{Conclusions}
\label{sec:conclusions}

%Even-odd effect in Kitaev multilayers. No magnetic order considered. 
In the present work, we have studied the magnetic properties of coupled 
Kitaev honeycomb multilayers. These systems are relevant to actual 
Kitaev materials since these actually consist on Kitaev honeycomb stacks. 
Also they are useful for the understanding of 
the Kitaev model on a honeycomb lattice with $S>1/2$ for which no exact solution exists. We have treated a model with an arbitrary number of 
coupled Kitaev layers 
stacked on top of each other using Abrikosov fermions. A key prediction 
from our mean-field decoupling without the magnetic channels
is that while a gapped QSL occurs 
in even-layered Kitaev models, odd-layered systems host 
ungapped QSLs. 

Previous numerical work \cite{khait_characterizing_2021,chen_excitation_2022} 
suggests the existence of a gapped QSL ground state in the $S=1$ 
2D Kitaev model. Analytical work\cite{ma_mathbbz_2_2023} indicates that a gapped QSL should be the ground state of the integer Kitaev model.
The multilayered Kitaev model can be mapped onto the 2D Kitaev model 
with integer (half-integer) $S$ for even (odd) FM coupled layers at sufficiently large $J$. Thus, the even-odd effect in the multilayered 
Kitaev model is in agreement with such previous 
numerical/analytical works on the spin-$S$ Kitaev model on a honeycomb lattice.

The PSG analysis helped us in constructing a fermionic mean-field {\it Ansatz} that is found to be the lowest energy one even on the bilayer system. Interestingly, for multilayer systems, an alternation of KSLs with opposite signs on all bonds from one layer to the other is stabilized, and is related by a specific inversion symmetry between the layers. One could have thought that the simple and direct reflection symmetry would have been the natural one instead. This peculiar {\it Ansatz} enforces the interlayer bonds to be strictly zero for the liquid solutions, as verified numerically by solving the self consistent equations.  

The validity of the mean-field and PSG approach is established by comparing with ED calculations.
Our mean-field energies deviate more strongly from the exact ED energies as $J$ increases. However, Gutzwiller projection of the mean-field {\it Ansatz} in the large-$J$ limit brings the energy dependence on $J$ in good agreement with the ED energies. While at $J=0$ the QSL {\it Ansatz} recovers the exact KSL solution, at large $J$, the Gutzwiller projected QSL coincides with the exact ground state. This indicates that the mean-field QSLs found are good candidates for the solution of the multilayered Kitaev model and can be considered adiabatically connected with the exact ED ground state.

Kitaev multilayers behave quite differently under the action of a magnetic field in the $[111]$ direction depending on the type of exchange coupling between the layers. The gapped KSL is stable in a
finite $B$-$J$ parameter region in which the critical
field to the FP phase, $B_c$, is rapidly suppressed by the action of a FM 
Heisenberg coupling independently of the number of Kitaev layers. In contrast,
$B_c$, increases with $J$ in AFM coupled Kitaev bilayers since $J$
counteracts the effect of the magnetic field. 

%Relevance to actual materials.

The $\alpha$-RuCl$_3$ material consists of stacks of honeycomb layers on top of each 
other \cite{janssen_magnon_2020} with a significant 
interlayer AFM coupling $J=1.3$ meV compared to the Kitaev in-plane coupling $-K=3-4$ meV. On the other hand the honeyomb iridate, H$_3$LiIr$_2$O$_6$, is a good candidate for realizing Kitaev physics \cite{kitagawa_spinorbital-entangled_2018}.
Estimates of the model parameters in
this material \cite{slagle_theory_2018} 
give: $-K \sim J =10 $ meV.
According to the phase diagrams of Fig.~\ref{fig:structure}, the $\alpha$-RuCl$_3$ behaves as a set of decoupled KSLs while in H$_3$LiIr$_2$O$_6$ ($-J/K \sim 1$) interlayer correlations may lead to a gapped (ungapped) QSL with interlayer correlations for odd (even) number of layers. Our study may be relevant to BaCo$_2$(AsO$_4$)$_2$ compounds on which small applied magnetic fields melt the magnetic order inducing a KSL\cite{zhong_weak-field_2020}.

%\textcolor{blue}{More examples of FM vs. AFM materials?}  

\begin{acknowledgments}
	We acknowledge financial support from (Grant No. PID2022-139995NB-I00) MINECO/FEDER, Uni\'on Europea, from the Mar\'ia de Maeztu Programme for Units of Excellence in R\&D 
	(Grant No. CEX2018-000805-M). AR acknowledges support from {\it Agence Nationale de la Recherche} (ANR), project FlatMoi (Grant No. ANR-21-CE30-0029).
\end{acknowledgments}

\appendix

\section{Projective symmetry group of the multilayer Kitaev liquids}
\label{app:psg}

\subsection{A reminder of the single layer Kitaev spin liquid}

In order to analyze the projective symmetry group (PSG) of the multilayer Kitaev spin liquids, we follow the construction and notations introduced in [\onlinecite{you_doping_2012}]. Let us here briefly recall their main results for the single layer before considering the multilayer case. The Abrikosov decomposition of the spin at site $i$ introduced in the main text as $S_i^a = \frac{1}{2} f_i^\dagger \sigma^a f_i$  with $f_i = ( f_{i \uparrow} , f_{i \downarrow} )$, the spinon operator $f_{i \sigma}$ possesses an additional SU(2) gauge structure that is easier to see by arranging the matrix as\cite{hermele_su2_2007}:
\begin{eqnarray}
F_i = \begin{bmatrix} f_{i \uparrow} & - f_{i \downarrow}^\dagger \\ f_{i \downarrow} &f_{i \uparrow}^\dagger \end{bmatrix}.
\end{eqnarray}
In function of this spinon matrix, the spin can be recast as 
\begin{eqnarray}
S_i^a = \frac{1}{4} \text{Tr} F_i^\dagger \sigma^a F_i.
\end{eqnarray}
It is easy to verify that any right SU(2) transformation $F_i \to F_i G : G \in $ SU(2) leaves the spin unchanged while any left SU(2) transformation  $U^\dagger F_i \to F_i: U^\dagger \in $ SU(2) is the usual spin rotation. Thus any operator acting on the spin will also affect the SU(2) gauge space, and it is the aim of the PSG analysis\cite{wen_quantum_2002} to specify how the symmetry operations of the model behave within this gauge freedom. 
Therefore, a symmetry operation consists to a spin operation $U_g$ belonging to a group operation $g \in$ SU(2), with a corresponding gauge operation $G_g$ in such a way that 
\begin{eqnarray}
F_i \to U_g^\dagger(i) F_{g(i)} G_g (i).
\end{eqnarray}

\begin{figure}[t!]
\centering
\includegraphics[width=0.3\textwidth,clip]{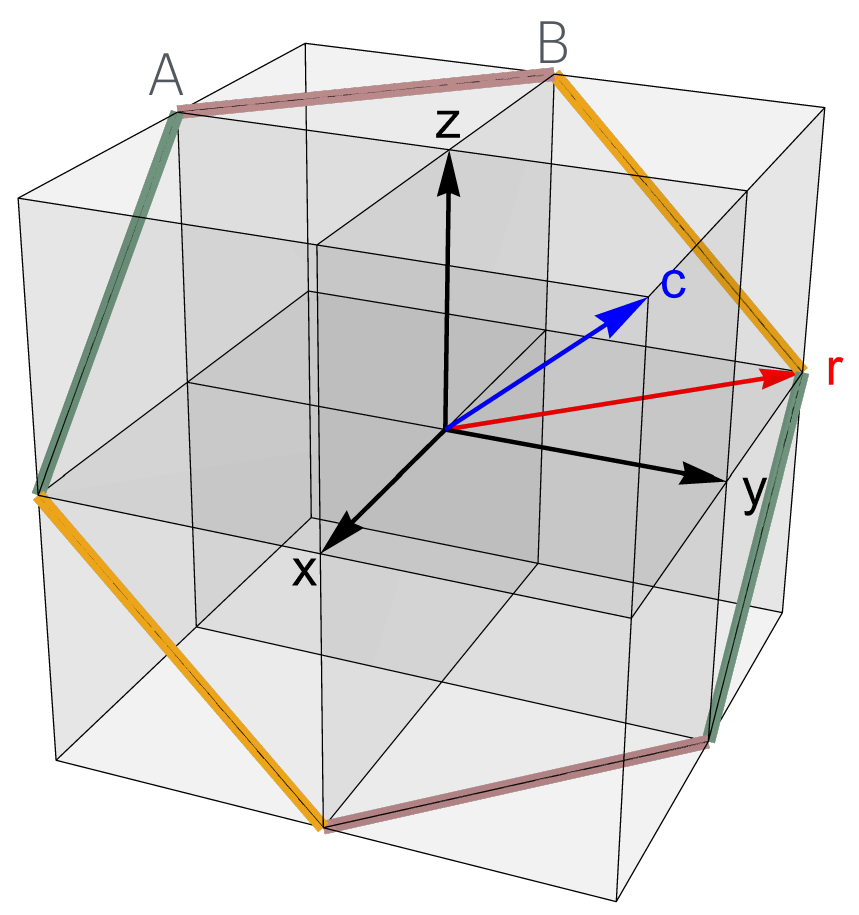} 
    \caption{Illustration of a hexagonal plaquette of the single layer Kitaev model with the three different bonds, $S^xS^x$ (yellow), $S^yS^y$ (green) and $S^zS^z$ (red).  The  $c$ vector (blue arrow) represents the axis of the $C_6$ sixfold rotation of the model.  The $\sigma$ reflection is across the $x=y$ plane oriented  along the $r$  vector (red arrow). }
\label{fig:FigApp00}
\end{figure}

As pointed out in [\onlinecite{wang_multinode_2020}],  a quantum spin liquid (QSL) ground state should preserve the full space group symmetry whose point group, for the single layer case, is $D_{3d} \times \mathbb{Z}_2^\tau$. So in addition to the two translations $T_1$ and $T_2$, we are left with three additional generators,  an operation of a sixfold rotation $C_6$ of axis ${\bf n}_{c} = ({\bf u}_x+{\bf u}_y+{\bf u}_z)/\sqrt{3}$ (see Fig.\ref{fig:FigApp00}) followed by a reflection across the lattice plane,  a reflection $\sigma$ across the $x=y$ plane of axis ${\bf n}_{\sigma} = ({\bf u}_y-{\bf u}_x)/\sqrt{2}$ , and an additional time-reversal symmetry $\tau$ of the Kitaev model, which is not site dependent and acts as $i \sigma_2$ followed by $K$ the complex conjugate operator flipping the sign of the imaginary unit, {\it i.e. } $K i \to -i$.
 
 As noticed in [\onlinecite{you_doping_2012}], the spin operations $U_g(i) = U_g$ are always site independent as well in the liquid state, so that the site index $i$ can be omitted. One can then obtain the expression of the symmetry operations of Fig.\ref{fig:FigApp00} and obtain their expressions for the two sub-lattice sites $A$ and $B$ as 
\begin{eqnarray}\label{eq1}
	U_{T_1} &=& U_{T_2} = 1, \nonumber \\
	U_\tau(A) &=&U_\tau(B) = i \sigma^y \nonumber \\
	U_{C_6} (A) &=& U_{C_6} (B) = \sigma_{C_6}, \\
	U_{\sigma} (A) &=& U_{\sigma} (B) = \sigma_{\sigma}, \nonumber
\end{eqnarray}
where we have defined 
\begin{eqnarray}
	\sigma_{C_6} &=& e^{ -i \frac{2 \pi}{3} {\bf n}_c . \boldsymbol{\sigma} } = \frac{1}{2} \left( \sigma_0 + i  \sum_{a=x,y,z}^3 \sigma^a \right), \\
	\sigma_{\sigma} &=& e^{ -i \pi {\bf n}_c . \boldsymbol{\sigma} } = \frac{i}{\sqrt{2}} ( \sigma^x - \sigma^y).
\end{eqnarray}

The standard way to proceed is to derive and solve all the algebraic relations of a given geometry and to obtain the possible physical {\it Ans\"atze} compatible with those symmetries and PSG representations (see for examples [\onlinecite{you_doping_2012,wang_spin-liquid_2006,bieri_gapless_2015,wang_schwinger_2010,reuther_classification_2014,wang_multinode_2020}] en references therein). In [\onlinecite{you_doping_2012}] though, the authors were searching to make the projective construction of  the Kitaev Spin liquid by considering additional constraints this specific  {\it Ansatz}  implies. 

The Kitaev model is solved exactly\cite{kitaev_anyons_2006} by introducing four Majorana fermions $\chi_i^\alpha$ ($\alpha = 0,x,y,z$) on each site of the lattice and by mapping the spin operators as $S_i^a = i \chi_i^0 \chi_i^a$. The Majorana operators act on a 4-dimensional Fock space whereas the Hilbert space of the spin is identified with a 2-dimensional subspace of this space. To project onto the physical subspace, the state has to obey the physical constraint $\chi_i^0\chi_i^x\chi_i^y\chi_i^z = 1/4$\cite{kitaev_anyons_2006}. Depending the given SU(2) gauge, those Majorana fermions have been shown to be related to the Abrikosov fermions $f_{i \sigma}$ by the simple relation\cite{burnell_su2_2011}
\begin{eqnarray}
F_i = \frac{1}{\sqrt{2}} ( \chi_i^0 \sigma_0 + i  \chi_i^x \sigma^x + i \chi_i^y \sigma^y +i  \chi_i^z \sigma^z )
\end{eqnarray}
corresponding to specific Majorana modes of the Abrikosov fermions $f_{i \uparrow} = ( \chi_i^0  + i \chi_i^z)/\sqrt{2}$ and $f_{i \downarrow} = ( i \chi_i^x  - \chi_i^y)/\sqrt{2}$. In terms of Majorana bilinears, the Hamiltonian reads
\begin{eqnarray}
H = J_K \sum_{\langle i,j\rangle_a} \left[ i u^a_{ij} \chi_i^0 \chi_j^0 + i u^0_{ij} \chi_i^a \chi_j^a  -  u^0_{ij} u^a_{ij} \right],
\end{eqnarray}
where we have defined the bond parameters $ u^\alpha_{ij} = \langle \frac{i}{2}  \chi_i^\alpha \chi_j^\alpha  \rangle$ ($\alpha=0,x,y,z$) is a mean field {\it Ansatz} that, after being solved self-consistently, gives the solution $u^0_{ij} =  -0.262433$ and $u^a_{ij} =  1/2$ for $a=x,y,z$. Note that changing the signs of all  bond parameters yields to a second compatible solution.  

As mentioned by You {\it et al.}\cite{you_doping_2012}, a key point in the projective construction of the Kitaev spin liquid is hat within the Majorana construction\cite{kitaev_anyons_2006}, $\chi^0$ is a special flavor which has not to be mixed with other flavors, such that any PSG operation should obey this constraint.  Noticing that $\chi^0$ appears in the $F$ spinon matrix as $\chi^0 \sigma_0$ and that as said $F$ transforms under a PSG operation as $F \to U_g^\dagger F G_g$, we thus have, up to a sign factor, $\chi^0 \sigma_0 \to \pm \chi^0 U_g^\dagger G_g$. It is thus sufficient to require $G_g = \pm U_g$ whatever is $g \in $ SG to preserve the flavor of $\chi^0$. Said in other words, one can enforce the gauge operations $G_g$ to always follow the spin operation $U_g$ up to a sign factor which can be chosen in order to keep the bond parameters $u^\alpha_{ij} = \langle \frac{i}{2} \chi_i^\alpha \chi_j^\alpha  \rangle$ unchanged under $C_6$, $\sigma$, and $\tau$ symmetries\cite{you_doping_2012}. Finally, one gets the gauge operations $G_g$ defined as 
\begin{eqnarray}
	\label{eq2}
	G_{T_1} &=& G_{T_2} = 1, \nonumber \\
	G_\tau(A) &=&-G_\tau(B) = i \sigma^y \nonumber \\
	G_{C_6} (A) &=& -G_{C_6} (B) = \sigma_{C_6}, \\
	G_{\sigma} (A) &=& -G_{\sigma} (B) = \sigma_{\sigma}. \nonumber
\end{eqnarray}
According the PSG classification of You {\it el al.}, the single layer Kitaev spin liquid belongs to the zero-flux class [I-B] PSG (see Appendix A of [\onlinecite{you_doping_2012}] for more details). 

\subsection{Case of the bilayer Kitaev spin liquid}

\begin{figure}[t!]
\centering
\includegraphics[width=0.4\textwidth,clip]{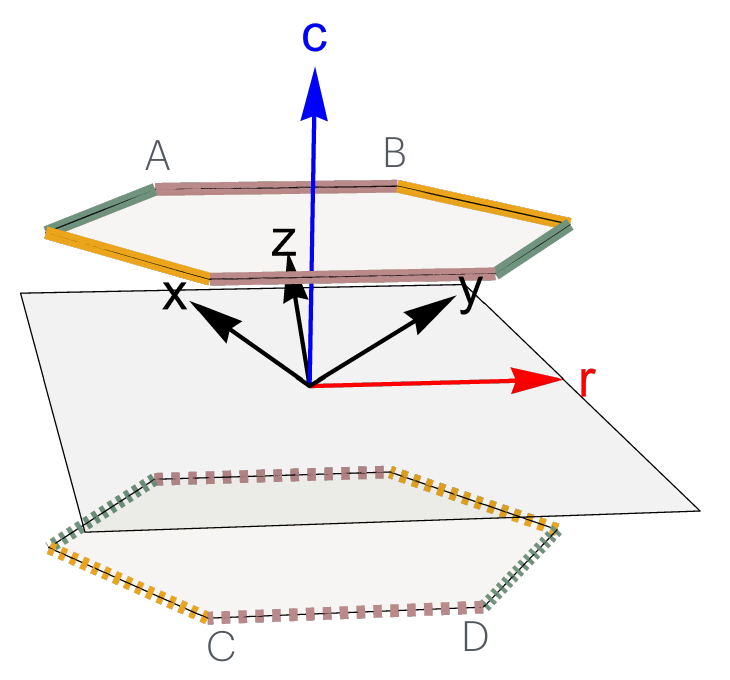} 
    \caption{Bilayer system with additional inversion plane. The three different bonds are represented in  $S^xS^x$ (yellow), $S^yS^y$ (green) and $S^zS^z$ (red), with continuous lines for the upper plane and dashed lines for the lower one.  $A, B$ sublattices belong to layer 1 and $C,D$ to layer 2. The  $c$ vector (blue arrow) represents the axis of the $C_6$ sixfold rotation of the model.  The $\sigma$ reflection is across the $x=y$ plane oriented  along the $r$  vector (red arrow). The additional symmetry consists in a $\pi$ rotation of same axis $r$ accompanied by a flip of the two layers around this axis in such a way that $A \to C$, $B \to D$ and $ x \to -x$, $y \to -y$.}
\label{fig:FigApp01}
\end{figure}
We define the bilayer system as in Fig.\ref{fig:FigApp01}, where one layer is on top of the other. It is not necessary to use 3-dimensional coordinates to describe the model. Instead, we introduce $C(D)$ sub-lattice site on the lower layer respectively corresponding to $A(B)$ on the upper layer.  The bilayer system also possesses additional interlayer bonds $(A,C)$ and $(B,D)$ that are coupled either ferromagnetically or antiferromagnetically.

We augment the PSG by starting from the solutions of the single layer case\cite{you_doping_2012} and adding the additional group stabilizers implied by the new lattice symmetry $C_r$ defined as:
\begin{eqnarray*}
	C_r: (x,y,A) &\to& (-x,-y, C), \\
	(x,y,B) &\to& (-x,-y, D),\\
	(x,y,C) &\to& (-x,-y, A),\\
	(x,y,D) &\to& (-x,-y, B).
\end{eqnarray*} 
Combined with the stabilizers of the single layer (see Appendix A of [\onlinecite{you_doping_2012}]), one get six additional relations for the symmetry group generated by our six generators $(T_1, T_2, \tau, C_6, \sigma, C_r)$
\begin{eqnarray}
	C_r^2 &=& 1, \\
	\tau C_r \tau C_r^{-1} &=& 1, \\
	\sigma C_r \sigma C_r &=& 1,\\
	C_6 C_r C_6 C_r^{-1} &=& 1,\\
	T_\alpha C_r T_\alpha C_r^{-1} &=& 1.
\end{eqnarray}
As said above, following You's notations\cite{you_doping_2012}, the single layer KSL belongs in PSG class [I-B] with corresponding ${\mathbb Z}_2$ variables
\begin{eqnarray}
\eta_1 = 1,~\eta_8 = -1,~\eta_1 = -1,~\eta_{11} = 1,\nonumber \\ ~\eta_{12} = 1,~\eta_{13} = -1,~\eta_{14} = -1.
\end{eqnarray}
With this definition, there is no site dependence of the symmetries and we can proceed with $C_r$ in order to define the PSG representation. Solving the algebraic equations of the PSG representations of our additional stabilizer conditions by keeping the KSL properties (independence of the site of any symmetry) yield to the modified symmetries and gauges: 
\begin{eqnarray}\label{eq3}
	U_{T_1} &=& U_{T_2} = 1,  \\
	U_\tau(A) &=&U_\tau(B) =U_\tau(C) =U_\tau(D) = i \sigma^y \nonumber \\
	U_{C_6} (A) &=& U_{C_6} (B) =U_{C_6} (C) =U_{C_6} (D) = \sigma_{C_6}, \nonumber \\
	U_{\sigma} (A) &=& U_{\sigma} (B) =U_{\sigma} (C) =U_{\sigma} (D) = \sigma_{\sigma}, \nonumber \\
	U_{r} (A) &=& U_{r} (B) =U_{r} (C) =U_{r} (D) = \sigma_{r}, \nonumber
\end{eqnarray}
and 
\begin{eqnarray}\label{eq3}
	G_{T_1} &=& G_{T_2} = 1,  \\
	G_\tau(A) &=&-G_\tau(B) = \epsilon_\tau G_\tau(C) =- \epsilon_\tau G_\tau(D) = i \sigma^y \nonumber \\
	G_{C_6} (A) &=& -G_{C_6} (B) =\epsilon_{C_6} G_{C_6} (C) =- \epsilon_{C_6} G_{C_6} (D) = \sigma_{C_6}, \nonumber \\
	G_{\sigma} (A) &=& G_{\sigma} (B) =\epsilon_{\sigma} G_{\sigma} (C) =-\epsilon_{\sigma} G_{\sigma} (D) = \sigma_{\sigma}, \nonumber \\
	G_{r} (A) &=& \epsilon_r G_{r} (B) =- G_{r} (C) =- \epsilon_r G_{r} (D) = \sigma_{r}, \nonumber
\end{eqnarray}
where we have introduced the new spin symmetry $U_r$ as defined in Fig.\ref{fig:FigApp01}, and ${\mathbb Z}_2$ variables $\epsilon_i$ accounting for the possibility of having one of the two KSL solution on the lower plane, once the one of the upper plane is fixed. Note that, as said, the symmetry $r$ is also performing a $\pi$-rotation of the lattice of axis $\vec{r}$, as displayed in Fig.\ref{fig:FigApp01}, but the {\it Ansatz} is site independent. 

We show in Table.\ref{tabletransform} how the Majorana spinons are transformed under these spin and gauge symmetries.
\begin{table}
	\begin{tabular}{p{0.07\textwidth}p{0.07\textwidth}p{0.07\textwidth}p{0.07\textwidth}p{0.07\textwidth}p{0.07\textwidth}}
	\hline
	\hline
	g~: &  $\tau$ &$T_\alpha$ & $C_6$ & $\sigma$  & $C_r$ \\
	\hline 
	$\chi_{A}^{0}~\to$ &$ \chi_{A}^{0}$ &$\chi_{A}^{0}$ &$\chi_{B}^{0}$ &$\chi_{B}^{0}$ &$\chi_{C}^{0}$  \\
	$\chi_{A}^{x}~\to$ &$ \chi_{A}^{x}$ &$\chi_{A}^{x}$ &$\chi_{B}^{z}$ &$-\chi_{B}^{y}$ &$-\chi_{C}^{y}$  \\
	$\chi_{A}^{y}~\to$ &$ \chi_{A}^{y}$ &$\chi_{A}^{y}$ &$\chi_{B}^{x}$ &$-\chi_{B}^{x}$ &$-\chi_{C}^{x}$  \\
	$\chi_{A}^{z}~\to$ &$ \chi_{A}^{z}$ &$\chi_{A}^{z}$ &$\chi_{B}^{y}$ &$-\chi_{B}^{z}$ &$-\chi_{C}^{z}$  \\
	$\chi_{B}^{0}~\to$ &$-\chi_{B}^{0}$ &$\chi_{B}^{0}$ &$-\chi_{A}^{0}$ &$-\chi_{A}^{0}$ &$-\chi_{D}^{0}$  \\
	$\chi_{B}^{x}~\to$ &$-\chi_{B}^{x}$ &$\chi_{B}^{x}$ &$-\chi_{A}^{z}$ &$\chi_{A}^{y}$ &$\chi_{D}^{y}$  \\
	$\chi_{B}^{y}~\to$ &$-\chi_{B}^{y}$ &$\chi_{B}^{y}$ &$-\chi_{A}^{x}$ &$\chi_{A}^{x}$ &$\chi_{D}^{x}$  \\
	$\chi_{B}^{z}~\to$ &$-\chi_{B}^{z}$ &$\chi_{B}^{z}$ &$-\chi_{A}^{y}$ &$\chi_{A}^{z}$ &$\chi_{D}^{z}$  \\
	$\chi_{C}^{0}~\to$ &$\epsilon_\tau \chi_{C}^{0}$ &$\chi_{C}^{0}$ &$ \chi_{D}^{0}$ &$\epsilon_{\sigma} \chi_{D}^{0}$ &$\epsilon_{C_r} \chi_{A}^{0}$  \\
	$\chi_{C}^{x}~\to$ &$\epsilon_\tau \chi_{C}^{x}$ &$\chi_{C}^{x}$ &$\chi_{D}^{z}$ &$-\epsilon_{\sigma} \chi_{D}^{y}$ &$-\epsilon_{C_r} \chi_{A}^{y}$  \\
	$\chi_{C}^{y}~\to$ &$\epsilon_\tau \chi_{C}^{y}$ &$\chi_{C}^{y}$ &$\chi_{D}^{x}$ &$-\epsilon_{\sigma} \chi_{D}^{x}$ &$-\epsilon_{C_r} \chi_{A}^{x}$  \\
	$\chi_{C}^{z}~\to$ &$\epsilon_\tau \chi_{C}^{z}$ &$\chi_{C}^{z}$ &$\chi_{D}^{y}$ &$-\epsilon_{\sigma} \chi_{D}^{z}$ &$-\epsilon_{C_r} \chi_{A}^{z}$  \\
	$\chi_{D}^{0}~\to$ &$-\epsilon_\tau \chi_{D}^{0}$ &$\chi_{D}^{0}$ &$-\chi_{C}^{0}$ &$-\epsilon_{\sigma} \chi_{C}^{0}$ &$-\epsilon_{C_r} \chi_{B}^{0}$  \\
	$\chi_{D}^{x}~\to$ &$-\epsilon_\tau \chi_{D}^{x}$ &$\chi_{D}^{x}$ &$-\chi_{C}^{z}$ &$\epsilon_{\sigma} \chi_{C}^{y}$ &$\epsilon_{C_r} \chi_{B}^{y}$  \\
	$\chi_{D}^{y}~\to$ &$-\epsilon_\tau \chi_{D}^{y}$ &$\chi_{D}^{y}$ &$-\chi_{C}^{x}$ &$\epsilon_{\sigma} \chi_{C}^{x}$ &$\epsilon_{C_r} \chi_{B}^{x}$  \\
	$\chi_{D}^{z}~\to$ &$-\epsilon_\tau \chi_{D}^{z}$ &$\chi_{D}^{z}$ &$-\chi_{C}^{y}$ &$\epsilon_{\sigma} \chi_{C}^{z}$ &$\epsilon_{C_r} \chi_{B}^{z}$ \\
	\hline
	\hline
	\end{tabular}
	\caption{Action of the PSG on the Majorana fermions of the bilayer system. Variables $\epsilon_i$ are independent ${\mathbb Z}_2$ variables taking values $\pm 1$.}
	\label{tabletransform}
\end{table}

\section{Gutzwiller projection at $K=0$.}
\label{app:largeJ}

Here we provide details about the states formed in the large-$J$ limit 
of the bilayer and trilayer Kitaev model.
To do so, we first consider the $K=0$ case, for which only the interlayer couplings are relevant. We then solve exactly the Heisenberg model which reduces for the bilayer and the trilayer case respectively to very simple 2-site and a 3-site problems. After performing the proper Gutzwiller projection, we compare our results to ED calculations.

\subsection{Case of the bilayer: dimers}

In the case of the bilayer, all sites of the system are paired up forming dimers between the two layers in such a way that the ground state is simply the tensorial product of a set of isolated and static dimers, induced by the AFM or FM coupling. 

The Heisenberg model on such a dimer can be expressed in terms of  
Abrikosov fermions using Eq. (\ref{eq:inter}) with $i=j, l=1, l'=2$ as:
\begin{eqnarray}
H &=& -J \hat{\bf S}_1  \hat{\bf S}_2 = -\hat{\bf h}^\dagger_{1,2} T\hat{\bf h}_{1,2} - \hat{\bf p}^\dagger_{1,2} T \hat{\bf p}_{1,2} 
\nonumber \\
&=& {J \over 8} (3 h^{0\dagger}_{1,2} h^0_{1,2} + 3 p^0_{1,2} p^0_{1,2} - h^{x\dagger}_{1,2} h^x_{1,2} - p^{x\dagger}_{1,2} p^x_{1,2} 
\nonumber \\
&-& h^{y\dagger}_{1,2} h^y_{1,2} - p^{y\dagger}_{1,2} p^y_{1,2} 
-h^{z\dagger}_{1,2} h^z_{1,2} - p^{z\dagger}_{1,2} p^z_{1,2} ),
\nonumber \\
\end{eqnarray}

{\it Antiferromagnetic, $J<0$, dimer.} 
In this case, we only keep $h_{12}^{0} \neq 0 \in \mathcal{R} $ so that the mean-field Hamiltonian reduces to:
\begin{equation}
\mathcal{H}_J={3 \over 8} J ( h_{12}^{0*} \hat{h}^0_{12} + {\hat h}_{12}^{0\dagger} h^0_{12} ).
\end{equation}
with $h^0_{12}= \langle f^\dagger_{1 \ua} f_{2 \ua} + f^\dagger_{1 \da} f_{2 \da} \rangle$. This implies solving:
\begin{equation}
\mathcal{H}_{J\sigma} | \psi^{MF}_\sigma \rangle = \epsilon_\sigma |\psi^{MF}_\sigma  \rangle,
\end{equation}
where:
\begin{equation}
\mathcal{H}_{J\sigma}= {3 \over 8} J \begin{pmatrix}
 0   &   h^{0*}_{12} \\
 h^{0}_{12} &  0    \\
\end{pmatrix}.
\end{equation}
The lowest energy eigenstate for each spin is: $|\psi_{0\sigma} \rangle = {1 \over \sqrt{2} }  (f^\dagger_{1 \ua} + f^\dagger_{2 \ua} ) | 0 \rangle$ with energy: $\epsilon_\sigma =- {3 \over 8} |J|$.  The mean-field energy of the two-site dimer reads:
\begin{equation}
E=\langle \mathcal{H}_J \rangle + {3 |J| \over 8}|h^{0}_{12}|^2=
-{3 |J| \over 4} h^{0}_{12}  + {3 |J| \over 8} |h^{0}_{12}|^2.
\end{equation}
with minimum energy:
\begin{equation}
E=- {3|J| \over 8}.
\label{eq:mfafmdi}
\end{equation}
The mean-field ground state reads:
\begin{eqnarray}
|\Psi_0^{MF} \rangle &=& | \psi_{0 \ua} \rangle  | \psi_{0 \da} \rangle  \\ &=& {1 \over 2} ( f^\dagger_{1\ua} f^\dagger_{1\da} + f^\dagger_{1\ua} f^\dagger_{2\da} - f^\dagger_{1\da} f^\dagger_{2\ua} +
f^\dagger_{2\ua} f^\dagger_{2\da}) | 0 \rangle. \nonumber 
\end{eqnarray}
The no-double occupancy constraint is automatically fulfilled on average due to particle-hole symmetry assuming $\lambda^a=0$.
By Gutzwiller projection of the mean-field state we find:
\begin{equation}
P_G | \Psi_0^{MF} \rangle \propto (f^\dagger_{1\ua} f^\dagger_{2\da} - f^\dagger_{1\da} f^\dagger_{2\ua}) | 0 \rangle,
\end{equation}
the expected exact singlet state with the ED energy: $E_{exact}=-3 J/4$.
Note that the energy of the mean-field {\it Ansatz} in Eq. (\ref{eq:mfafmdi}) is half the exact energy. 

{\it Ferromagnetic, $J>0$, dimer.} Here we assume that only the pairing variational parameter: $p^z_{1,2}=-\langle  f_{1\ua} f_{2 \da} +f_{1\da} f_{2 \ua} \rangle$ is non-zero in order to approximate the full calculation. This 
leads to the mean-field Hamiltonian:
\begin{equation}
\mathcal{H}_J=-{J \over 8} ( p_{12}^{z*} \hat{p}^{z}_{12} + {\hat p}_{12}^{z\dagger} p^{z}_{12} ).
\end{equation}
which leads to the four-fold degenerate eigenenergies:
\begin{eqnarray}
\epsilon_{0\sigma}=-\Delta {J \over 16}  
\nonumber \\
\epsilon_{1\sigma}=\Delta {J \over 16}  
\end{eqnarray}

So the total mean-field energy reads:
\begin{equation}
E=-{J \over 4} \Delta + {J\over 8} \Delta^2,
\end{equation}
which attains its minimum value:
\begin{equation}
E = - {J \over 8}.
\label{eq:mffmdi}
\end{equation}

The mean-field ground state for this non-zero pairing amplitude reads:
\begin{equation}
    |\Psi_0^{MF} \rangle = (-f^\dagger_{1\ua} f^\dagger_{1\da} f^\dagger_{2\ua} f^\dagger_{2\da} - f^\dagger_{1\da} f^\dagger_{2 \ua} 
    - f^\dagger_{1\ua} f^\dagger_{2 \da} + f_{1\da} f_{2\da} f^\dagger_{1\da} f^\dagger_{2 \da} ) | 0 \rangle. 
\end{equation}
This BCS-type wavefunction also satisfies the no-double occupancy constraint on average. The Gutzwiller projected wavefunction would read: 
\begin{equation}
|\Psi_0 \rangle = P_G |\Psi_0^{MF} \rangle \propto (f^\dagger_{1\da} f^\dagger_{2 \ua} + f^\dagger_{1\ua} f^\dagger_{2\da} ) |0 \rangle
\end{equation}
recovering the expected triplet, $|S=1, S^z=0 \rangle$ state of the dimer
with exact energy: $E_{exact}=-{J \over 4}$ which is twice the mean-field energy in Eq. (\ref{eq:mffmdi}).

The two other $|S=1,S^z=\pm 1\rangle$ triplet states may be obtained allowing, for instance, for non-zero $p^x_{12}, p^y_{12}$ SC pairing amplitudes. 

\subsection{Case of the trilayer: trimers}

In the case of the trilayer, the $K=0$ limit reduces the Hamiltonian to a 3-site Heisenberg model on the 3 layers. Similarly to the bilayer case, the corresponding Heisenberg Hamiltonian can be written in terms of Abrikosov fermions as:
\begin{eqnarray}
H_J &=& -J (\hat{\bf S}_1  \hat{\bf S}_2 +\hat{\bf S}_2  \hat{\bf S}_3)
\nonumber \\
&=& -J (\hat{\bf h}^\dagger_{1,2} T\hat{\bf h}_{1,2} + \hat{\bf p}^\dagger_{1,2} T \hat{\bf p}_{1,2} )
\nonumber \\
&-& J (\hat{\bf h}^\dagger_{2,3} T\hat{\bf h}_{2,3} + \hat{\bf p}^\dagger_{2,3} T \hat{\bf p}_{2,3} )
\nonumber \\
&=& {J \over 8} (3 h^{0\dagger}_{1,2} h^0_{1,2} + 3 p^0_{1,2} p^0_{1,2} - h^{x\dagger}_{1,2} h^x_{1,2} - p^{x\dagger}_{1,2} p^x_{1,2} 
\nonumber \\
&-& h^{y\dagger}_{1,2} h^y_{1,2} - p^{y\dagger}_{1,2} p^y_{1,2} 
-h^{z\dagger}_{1,2} h^z_{1,2} - p^{z\dagger}_{1,2} p^z_{1,2} ),
\nonumber \\
&+&{J \over 8} (3 h^{0\dagger}_{2,3} h^0_{2,3} + 3 p^0_{2,3} p^0_{2,3} - h^{x\dagger}_{2,3} h^x_{2,3} - p^{x\dagger}_{2,3} p^x_{2,3} 
\nonumber \\
&-& h^{y\dagger}_{2,3} h^y_{2,3} - p^{y\dagger}_{2,3} p^y_{2,3} 
-h^{z\dagger}_{2,3} h^z_{2,3} - p^{z\dagger}_{2,3} p^z_{2,3} ).
\end{eqnarray}
Once again, this Hamiltonian is decoupled at the mean field level and solved for the AFM and FM cases.

{\it Antiferromagnetic, $J<0$, trimer.}

In this case, the mean-field Hamiltonian reads:
\begin{equation}
\mathcal{H}_{J\sigma}= {3 J \over 8} \begin{pmatrix}
 0   &  h^{0}_{12}  & 0 \\
 h^{0*}_{12} &  0  & h^{0}_{23}  \\
 0 &  h^{0*}_{23} & 0  \\
\end{pmatrix}  
\end{equation}
Assuming $h^{0}_{12}=h^{0}_{23}$, the eigenspectrum reads: 
\begin{eqnarray}
\epsilon_0 &=&-\sqrt{2} {3 \over 8} |J| h^{0}_{12}  
\nonumber \\
\epsilon_1 &=&0 
\nonumber \\
\epsilon_2 &=& \sqrt{2} {3 \over 8} |J| h^{0}_{12} 
\end{eqnarray}
with the corresponding eigenstates: $| \psi_{0\sigma} \rangle = {1 \over 2} ( f^\dagger_{1 \sigma} + \sqrt{2} f^\dagger_{2 \sigma} + f^\dagger_{3 \sigma}) |0 \rangle $, $| \psi_{1\sigma} \rangle = {1 \over 2} ( f^\dagger_{1 \sigma} + \sqrt{2} f^\dagger_{2 \sigma} + f^\dagger_{3 \sigma}) |0 \rangle $. 

The mean-field energy reads:
\begin{equation}
E=-{3 |J| \sqrt{2} \over 4} h^0_{12} + {3 | J| \over 4} |h_{12}^0|^2,
\end{equation}
with a minimum value:
\begin{equation}
E=-{3 |J| \over 8}.
\label{eq:mfafmtri}
\end{equation}

The $S^z=1/2$, mean-field ground state reads: 
\begin{eqnarray}
|\Psi_0^{MF} \rangle &=& | \psi_{0\ua} \rangle   | \psi_{0\da}  \rangle  | \psi_{1\ua} \rangle 
\nonumber \\
&=& {1 \over 4 \sqrt{2}} (-f^\dagger_{1 \ua}  f^\dagger_{1 \da} f^\dagger_{3 \ua} -\sqrt{2} f^\dagger_{1 \ua}  f^\dagger_{2 \da} f^\dagger_{3 \ua}- f^\dagger_{1 \ua}  f^\dagger_{3 \da} f^\dagger_{3 \ua}
\nonumber \\
&+& \sqrt{2} f^\dagger_{2 \ua}  f^\dagger_{2 \da} f^\dagger_{3 \ua} - \sqrt{2} f^\dagger_{2 \ua}  f^\dagger_{1 \da} f^\dagger_{3 \ua} + 2 f^\dagger_{2 \ua}  f^\dagger_{2 \da} f^\dagger_{1 \ua} 
\nonumber \\
&-& 2 f^\dagger_{2 \ua}  f^\dagger_{2 \da} f^\dagger_{3 \ua} +\sqrt{2} f^\dagger_{2 \ua}  f^\dagger_{3 \da} f^\dagger_{1 \ua}- \sqrt{2} f^\dagger_{2 \ua}  f^\dagger_{3 \da} f^\dagger_{3 \ua} 
\nonumber \\
&+& f^\dagger_{3 \ua}  f^\dagger_{1 \da} f^\dagger_{1 \ua} + \sqrt{2} f^\dagger_{3 \ua}  f^\dagger_{2 \da} f^\dagger_{1 \ua} + f^\dagger_{3 \ua}  f^\dagger_{3 \da} f^\dagger_{1 \ua}) |0 \rangle.
\nonumber \\
\end{eqnarray}
Note that the Gutzwiller projection leads to:
\begin{equation}
P_G | \Psi_0^{MF} \rangle  \propto ( f^\dagger_{1 \ua} f^\dagger_{2 \ua} f^\dagger_{3 \da} -2 
f^\dagger_{1 \ua} f^\dagger_{2 \da} f^\dagger_{3 \ua} + f^\dagger_{1 \da} f^\dagger_{2 \ua} f^\dagger_{3 \ua}) | 0 \rangle,
\end{equation}
recovering the exact $S^z=1/2$ ground state of the trimer. The $S^z=-1/2$ could be obtained in a similar way applying the projection on the corresponding $ S^z=-1/2$,  {\it i. e.}, $|\Psi_0^{MF} \rangle = | \psi_{0\ua} \rangle   | \psi_{0\da}  \rangle  | \psi_{1\da} \rangle $  mean-field state. The exact energy of this $S^z = \pm {1 \over 2} $ doublet is: $E_{exact}=-|J|$, much lower than the mean-field value in Eq. (\ref{eq:mfafmtri}).

{\it Ferromagnetic, $J>0$, trimer.}

As in the case of the dimers we assume that only the pairing $p_{1,2}^z$ amplitude is non zero emulating the full mean-field calculations of the main text. Hence, the mean-field Hamiltonian reads: 
\begin{equation}
\mathcal{H}_J=-{J \over 8} ( p_{12}^{z*} \hat{p}^{z}_{12} + {\hat p}_{12}^{z\dagger} p^{z}_{12} + p_{23}^{z*} \hat{p}^{z}_{23} + {\hat p}_{23}^{z\dagger} p^{z}_{23}),
\end{equation}
with the spectrum formed by the four-fold degenerate eigenenergies:
\begin{eqnarray}
\epsilon_{0\sigma}=-\Delta {J \over 8 \sqrt{2}}   
\nonumber \\
\epsilon_{1\sigma}= 0 
\nonumber \\
\epsilon_{2\sigma}=\Delta {J \over 8 \sqrt{2}},  
\end{eqnarray}
where we have assumed: $p_{12}^z=p_{23}^z=\Delta$.
This leads to the total mean-field energy:
\begin{equation}
E = -{ J \over  \sqrt{8}} \Delta +  {J \over 4} \Delta^2, 
\end{equation}
with minimum energy:
\begin{equation}
E= -{J \over 8}.
\label{eq:mffmtr}
\end{equation}
which is higher than the exact ground state energy:
$E^{MF}=-{ J \over 2 }$.
Again, the Gutzwiller projected wavefunction recovers the, say, 
$S^z= 1/2$, exact ground state wavefunction:
\begin{equation}
P_G | \Psi_0^{MF} \rangle  \propto ( f^\dagger_{1 \ua} f^\dagger_{2 \ua} f^\dagger_{3 \da} +2 
f^\dagger_{1 \ua} f^\dagger_{2 \da} f^\dagger_{3 \ua} + f^\dagger_{1 \da} f^\dagger_{2 \ua} f^\dagger_{3 \ua}) | 0 \rangle.
\end{equation}
of the FM trimer. The exact energy of the $S^z=\pm {1 \over 2}$ doublet 
is $E_{exact}=-{J\over 2} $ which is four times the mean-field energy in Eq. (\ref{eq:mffmtr}).
\bibliography{library}
 \end{document}